\documentclass[aps,pra,twocolumn]{revtex4-2}

\usepackage{graphicx} 
\usepackage{physics}
\usepackage{amsmath}
\usepackage{subcaption} 
\usepackage{amssymb}
\usepackage{hyperref}
\usepackage{xcolor}    
\usepackage{ragged2e}
\usepackage{microtype} 
\usepackage{tikz}                    
\usepackage{dblfloatfix} 
\usepackage{booktabs}
\usepackage{makecell}

\begin{document}

\title{Network-based prediction of drug combinations with quantum annealing}

\author{Diogo Ramos}
\affiliation{Centro de Física das Universidades do Minho e do Porto, LaPMET, Departamento de Física e Astronomia, Faculdade de Ciência, Universidade do Porto, Rua do Campo Alegre, Porto, 4169-007, Portugal}

\author{Bruno Coutinho}
\affiliation{Institute of Communications and Navigation, German Aerospace Center (DLR), Weßling, Germany}
\affiliation{Instituto de Telecomunicações (IT), Lisboa, Portugal}

\author{Duarte Magano}
\email[Corresponding author address: ]{duarte.magano@fc.up.pt}

\affiliation{Centro de Física das Universidades do Minho e do Porto, LaPMET, Departamento de Física e Astronomia, Faculdade de Ciência, Universidade do Porto, Rua do Campo Alegre, Porto, 4169-007, Portugal}
\affiliation{21strategies GmbH, Hallbergmoos, Germany}
\date{\today}

\begin{abstract}
The systematic discovery of effective drug combinations is a challenging problem in modern pharmacology, driven by the combinatorial growth of potential pairings and dosage configurations. 
Network medicine, modeling diseases and drugs as interconnected modules of the human protein-protein interactome, has emerged as a new paradigm for understanding disease mechanisms and drug action.
In this work, we propose a quantum annealing-based algorithm for identifying effective drug combinations.
Underlying our approach is the biologically motivated principle of `Complementary Exposure', which posits that therapeutic drug combinations target distinct yet complementary regions of a disease module. We translate this into a quadratic unconstrained binary optimisation problem.
We test our method for Diabetes Mellitus, Rheumatoid Arthritis, Asthma, and Brain Neoplasms, relying on experimentally validaded drug combinations for these diseases.
Our simulated quantum annealing experiments reveal that low-energy configurations align with biologically plausible combinations which may serve as prioritized candidates for subsequent experimental validation.
\end{abstract}

\maketitle

\onecolumngrid 

\vspace{1mm}
\noindent\textbf{Keywords:} quantum annealing; network medicine; drug combinations; complementary exposure

\vspace{4mm}

\twocolumngrid 
 
\allowdisplaybreaks[1]
\parindent0mm

\section{Introduction}

In modern medicine, therapies frequently combine multiple drugs to treat the same disease, since combination therapy has been shown to allow lower dosages and have fewer side effects~\cite{Xiaochen}. With over 1{,}000 drugs approved by the U.S. Food and Drug Administration targeting 3{,}000 human diseases \cite{Law2014}, finding the optimal drug combination is a complex process. 
The challenge is further amplified by the difficulty of predicting interactions among medications. Although physicians frequently rely on experience and intuition, addressing this problem ultimately requires a systematic approach.
It is well established that drugs act by targeting specific proteins in our cells, but knowing which proteins each drug targets is not sufficient.
Proteins operate within a complex network of interactions, the interactome, which encodes the mechanisms underlying cellular function.
While the full interactome remains far from mapped, we do have partial access to one of its key subnetworks: the protein-protein interaction (PPI) network. In this network, nodes represent proteins, and edges indicate known interactions.
To date, researchers have mapped over 243{,}000 interactions involving more than 16{,}000 proteins. Although it is difficult to estimate how close this is to the complete underlying network, the existing data has already yielded major insights into the inner workings of a cell. 
Notably, \cite{Barabasi2011,Menche2015} found that proteins associated with the same disease tend to be in the same neighbourhood of the PPI network, forming the so-called disease modules. Building on this intuition, Cheng \textit{et al.}~\cite{Cheng2019} proposed a systematic framework for predicting drug combinations. They introduced two main concepts: complementary exposure and overlapping exposure. 
The idea is that individual drugs are effective when they target proteins close to the corresponding disease modules, thereby maximising exposure. 
However, efficient drug-drug combinations require complementary action, meaning that each drug acts on a distinct subset of proteins within the module. 
In contrast, when multiple drugs target overlapping protein sets, the risk of side effects increases~\cite{Cheng2019}.

Quantum computation has already been proposed for a range of medical applications. In 2019, Grimsley \textit{et al.} introduced a promising approach that uses quantum adaptive variational simulations of molecular interactions for drug discovery~\cite{Grimsley2019}. In 2021, quantum computation was also applied to the optimisation of radiotherapy treatment plans~\cite{Niraula2021-et}. More recently, in 2023, Moutinho \textit{et al.} studied a quantum algorithm with direct medical implications that explicitly incorporates network structure: they proposed a quantum-walk-based method to identify missing links in complex networks, with a strong emphasis on biological networks such as PPI networks~\cite{Moutinho2023}. Later work suggested that this approach may offer a potential polynomial or exponential speed-up, depending on the network's structure~\cite {Moutinho2024,Duarte2023}. Once again using information from PPI networks, in 2025 Marques \textit{et al.}~\cite{marques2025} developed an algorithm for identifying potential cancer-related targets, achieving an exponential speed-up per prediction.

Beyond structural network mapping, Quantum Machine Learning (QML) and quantum-inspired frameworks are increasingly being deployed for complex biological classification tasks, ranging from the recognition of botanical leaf profile features~\cite{Kumar2022Quantum} to high-dimensional genomic sequence analysis~\cite{11231118} and the evaluation of plasma amino acid profiles for clinical diagnostics~\cite{10.1093/clinchem/hvaf192}. 
On the classical machine learning side, Physics-Informed Neural Networks (PINNs) offer a way to embed established physical laws into the network architecture, having been used to evaluate respiratory inhalation impedance~\cite{KUMAR2023107421}, simulate patient-specific 3D airflows~\cite{fluids10120306} and reconstruct the hidden transmission parameters of viral outbreaks from incomplete hospital data~\cite{HAN2024106671}.

Quantum annealing, a quantum computing technique tailored for solving combinatorial optimisation problems, also offers a promising direction \cite{farhi2000, AnnealingReview}. Unlike classical heuristics, which can struggle in high-dimensional energy landscapes, quantum annealers leverage quantum effects such as tunnelling to explore solution spaces differently \cite{das2008}. By approximately evolving the system toward low-energy states, they may find high-quality solutions more efficiently for certain problem types, though their practical advantage over classical methods remains an active area of research \cite{AnnealingScience,kim2025}. This capability aligns with the challenges of drug combination discovery, where evaluating every possible combination of two or more drugs can become computationally prohibitive \cite{salloum2024}. 

This emerging technology has been applied to a variety of state-of-the-art combinatorial and sampling problems. Notably, large-scale scheduling and routing \cite{lucas2014, PerezArmas2024}, portfolio optimisation \cite{sakuler2025portfolio,math12091291}, energy-grid resource allocation \cite{Blenninger2024QGRID} and materials design \cite{PhysRevResearch.2.013319}, since many such tasks map naturally to Quadratic Unconstrained Binary Optimisation (QUBO) formulations and can exploit specialised annealer embeddings and architectures. 
In medicine, quantum annealing has been proposed for tasks that impact drug discovery and molecular design, including conformation search and generation \cite{perdomo2012, Li2024}, peptide and protein design \cite{tucs2023, meuser2025, irback2024} and molecular docking \cite{Li2024docking}.
While largely exploratory today, these directions highlight promising hybrid workflows that combine classical preprocessing, network-based models, and annealing hardware for scalable combinatorial exploration in biomedical pipelines.

In this work, we propose a novel quantum annealing-based approach to identify potential drug combinations. Our algorithm builds on the `Complementary Exposure' principle of Cheng \textit{et al.}~\cite{Cheng2019}. We first express that principle as a QUBO problem suitable for implementation on current quantum annealers and then add biologically motivated constraints to better accommodate validated drug combination datasets. Finally, we discuss and demonstrate how the presented pipeline fits within the quantum annealing paradigm by employing a Simulated Quantum Annealing (SQA) sampler to extract meaningful drug-combination predictions.

We test our approach against Diabetes Mellitus, Rheumatoid Arthritis, Asthma, and Brain Neoplasms, and their respective validated drug combinations, discussing the algorithm's broader applicability in medicine.
Our framework  can generate recommendations for biologically plausible drug combinations, which may serve as prioritized candidates for subsequent experimental validation.

\section{Network Proximity Measures}

\begin{figure}[t]
\centering
\includegraphics[width=\columnwidth]{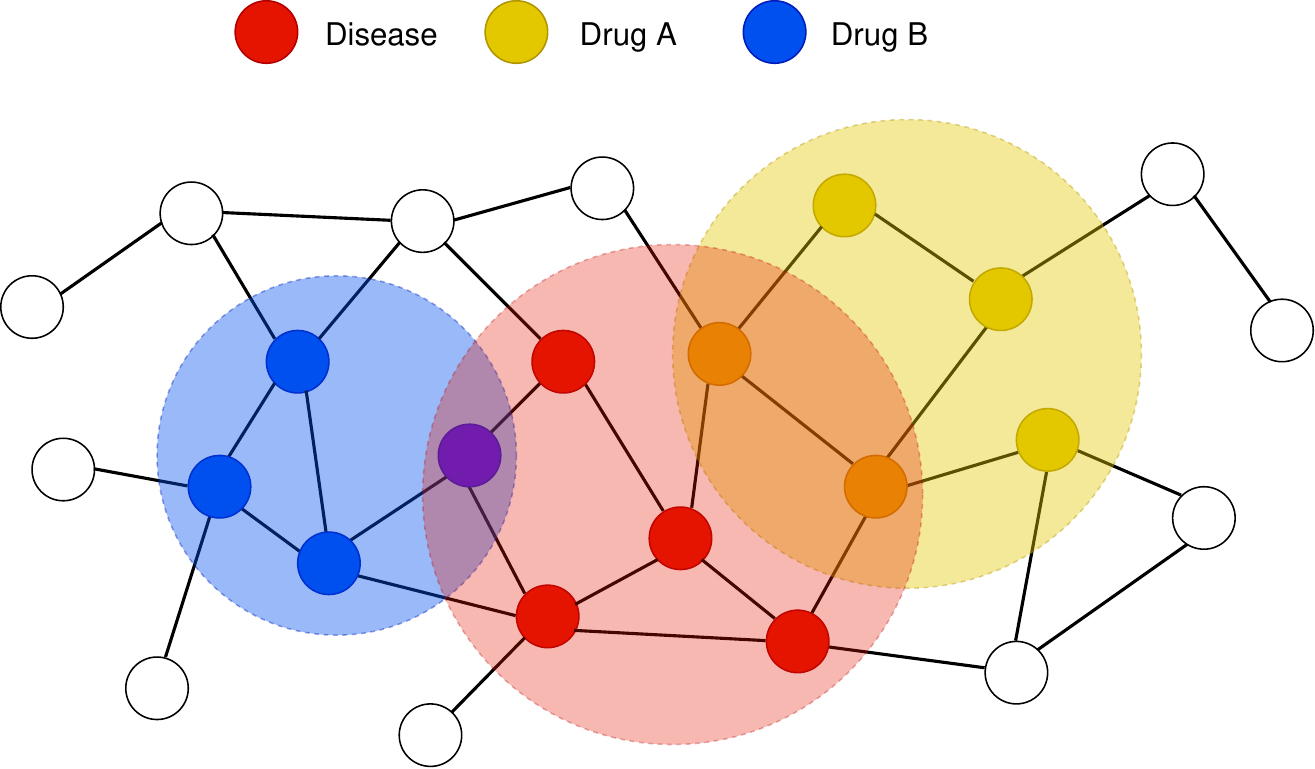}
\caption{ \justifying \textbf{Schematic of the `Complementary Exposure' principle.} Two drugs, A (yellow) and B (blue), represented by their protein targets, exhibit shared nodes with the disease module (red) in the human protein-protein interactome graph. Both drug node sets target distinct, non-overlapping regions of the disease module (red). This leads to effective perturbation of the disease-associated network neighborhood as quantified by the network proximity measures $s(A,B)\geq0$ and $z_A,z_B<0$.
}
\label{fig:CompExp}
\end{figure}

We represent molecular interactions using the human protein--protein interactome.
This is an undirected graph $\mathcal{G}=(V,E)$, where nodes $v\in V$ correspond to proteins and edges $(u,v)\in E$ designate experimentally supported physical interactions.
The network distances refer to shortest-path lengths on $\mathcal{G}$. Disease modules are defined as sets of disease-associated proteins $Y \subset V$. In practice, $Y$ is constructed by mapping curated disease-gene associations onto $\mathcal{G}$ and retaining only those whose proteins appear as nodes in the interactome, with genes absent from $\mathcal{G}$ being excluded since no network distance can be computed for them. However, the module need not form a single connected component. See Appendix~\ref{sec:datasets} for further details on the curated datasets.

The `Complementary Exposure' principle assumes that the therapeutic value of a drug combination depends on its ability to perturb the disease-associated network neighbourhood (the disease module) in a complementary manner, while avoiding redundant targeting or excessive overlap that can increase toxicity. A diagrammatic illustration of this relationship for a drug-drug-disease combination is provided in Figure~\ref{fig:CompExp}.

Furthermore, Cheng \textit{et al.}~\cite{Cheng2019} propose that network-based proximity can be used as a proxy to infer both the impact of a drug on a disease module and the extent of overlap between drugs. Namely, drug targets and disease modules within the human protein-protein interactome. 
Here, we refer to the two network-based metrics central to the `Complementary Exposure' principle: \textit{z-score} \cite{Guney2016} and the \textit{separation measure} \cite{Menche2015}.

The \textit{z-score} evaluates the network-based proximity between a drug $X$ and a disease module $Y$. It is defined as
\begin{equation}
    z(X,Y) = \frac{d(X,Y) - \mu_{(X,Y)}}{\sigma_{(X,Y)}}
\end{equation}
where $\mu_{(X,Y)}$ and $\sigma_{(X,Y)}$ represent the mean and standard deviation of distances from a null model. This model consists of an ensemble of degree-matched random node sets, which controls for degree-related topological biases \cite{Guney2016,Menche2015}. The distance $d(X,Y)$ between a set of drug targets $X$ 
and a set of disease proteins $Y$ is defined as the average shortest-path length from each disease protein $y\in Y$ to its nearest drug target $x\in X$:
\begin{equation} 
    d(X,Y)= \frac{1}{|Y|}\sum_{y\in Y}\min_{x\in X}d(x,y)
\end{equation}
A negative $z$-score indicates that drug targets are closer to the disease module than expected by chance, suggesting therapeutic relevance \cite{Guney2016}. 
Therefore, we give preference to disease-drug pairs with negative $z$-scores. This choice is consistent with the hypothesis that effective drugs localize near disease modules, as disease-associated genes cluster in interconnected neighborhoods \cite{Menche2015}.

While the $z$-score identifies drug-disease relationships and has been used to successfully predict monotherapies \cite{Guney2016}, its utility diminishes for combinations due to small drug-target sets and non-Gaussian randomization biases \cite{Cheng2019}. This phenomena is addressed by the \textit{separation measure} \cite{Menche2015}:
\begin{equation}
    s(A,B) = \langle d_{AB} \rangle - \frac{\langle d_{AA} \rangle + \langle d_{BB} \rangle}{2}
\end{equation}
where $\langle d_{AB} \rangle$ is the mean shortest path between targets of drugs $A$ and $B$, while $\langle d_{AA} \rangle$ and $\langle d_{BB} \rangle$ are the mean shortest distances of the targets of each drug individually. If a protein is a target of both sets, it has distance zero by definition.
The separation measure $s(A,B)$ serves as a topological indicator of drug interaction mechanisms. A negative value $s(A,B) < 0$ indicates overlapping drug targets within shared network neighborhoods. Conversely, a non-negative value $s(A,B) \geq 0$ indicates target localization in distinct topological regions, potentially leading to complementary therapeutic effects.  
Therefore, according to the `Complementary Exposure' principle, effective drug combination must satisfy two criteria: (i) its drugs exhibit significant proximity to the disease module, as evidenced by $z_{A} < 0$ and $z_{B} < 0$; and (ii) their targets occupy disjoint regions of the interactome, quantified by $s(A,B) \geq 0$. These criteria ensure that combinations simultaneously maximize disease module perturbation while minimizing redundant target engagement, a hallmark of clinically viable complementary therapies \cite{Cheng2019}.

Previous studies have corroborated these concepts by successfully predicting therapeutic drug combinations using models grounded in these network-based metrics \cite{AZAD2021, Cao2024}, and in one instance, experimentally validating the results \cite{iida2024transomics}.

\section{Quantum Annealing}  
\label{sec:qa_agc}  

Quantum annealing (QA) is a quantum computing paradigm specially suited for solving combinatorial optimisation problems \cite{AnnealingReview}. 
In QA a spin system evolves under a time-dependent Hamiltonian $H(t)$ \cite{QAIsing},
\begin{equation}  
H(t) = \left(1 - \frac{t}{T} \right) H_0 + \frac{t}{T} H_p.
\label{eq:homotopy}
\end{equation}  
QA carries a transition between an initial Hamiltonian $H_0$ at time $t=0$ and the \emph{problem} Hamiltonian at time $t=T$.
A typical choice for the initial Hamiltonian is
\[
H_0 = -\sum_i \sigma_i^x,
\]
since the ground state, a uniform superposition of all states in the computational basis, is simple to prepare in most platforms.
The problem Hamiltonian, 
\begin{equation}  
H_p = \sum_{i} h_i \sigma_i^z + \sum_{i<j} J_{ij} \sigma_i^z \sigma_j^z, 
\end{equation} 
encodes the cost function of the underlying optimisation problem such that its solution is found in the ground state of $H_p$.

The quantum adiabatic theorem ensures that, if the transition is slow enough (\textit{i.e.}, if $T$ is large enough), the system remains in the instantaneous ground state of $H(t)$ throughout the entire evolution \cite{farhi2000}.
It is typically stated as follows. 
Let $E_i(t)$ ($\ket{E_i(t)}$) be the $i$-th instantaneous eigenenergy (eigenstate) of $H(t)$; if the system is initialized in the ground state $\ket{E_0(0)}$ of $H_0$, it will remain (approximately) in the instantaneous ground state of $H(t)$ for all time if the following condition is satisfied \cite{Amin}
\begin{equation}  
1 \gtrsim \frac{\max_{s} \left| \langle E_i(s) | \partial_t H(t) | E_0(s) \rangle \right|}{\vert E_i - E_0 \vert^2}, \text{ for all } i \neq 0.
\label{eq:time_scaling}
\end{equation} 
A more rigorous  statement can be found in \cite{AmbainisRegev}.

These theoretical guarantees assume a perfect annealing machine, isolated from any environment interference. 
By contrast, real machines are subject to noise, which increases the probability of the system escaping the ground state. 
They should be considered heuristic solvers, requiring empirical performance analysis \cite{DWave}.

\section{Problem Hamiltonian}

While quantum annealing hardware natively operates on the Ising spin model, combinatorial optimization problems are often most intuitively expressed using binary decision variables. A QUBO problem is defined over a vector of binary decision variables $\mathbf{S} = (S_1, S_2, \ldots, S_n)$, $S_i \in \{0,1\}$, and takes the general form
\begin{equation}
    \min_{\mathbf{S}} \; \mathbf{S}^\top Q \, \mathbf{S} = \min_{\mathbf{S}} \left( \sum_{i} Q_{ii} \, S_i + \sum_{i<j} Q_{ij} \, S_i S_j \right),
    \label{eq:qubo_general}
\end{equation}
where $Q$ is a real-valued upper-triangular matrix. The diagonal coefficients $Q_{ii}$ encode the cost or benefit of selecting each individual variable, while the off-diagonal coefficients $Q_{ij}$ capture dependencies or interactions between pairs of variables. This formulation is equivalent to an Ising spin Hamiltonian via the transformation $S_i = (1 + \sigma^z_i)/2$, so that the ground state of the resulting Hamiltonian corresponds to the optimal QUBO solution.

Now, let $Y$ be a disease and \( D = \{X_1, X_2, \ldots, X_n\} \) a set of \( n \) candidate drugs.
We assign a binary decision variable $S_i \in \{0,1\}$ to each drug $X_i$, indicating its inclusion ($S_i = 1$) or exclusion ($S_i = 0$) from the combination. 
Following the Complementary Exposure principle each selected drug should exhibit significant proximity to the disease module while the targets of co-selected drugs should occupy distinct regions of the interactome. This translates nicely into a QUBO formulation, where the linear terms encode individual drug-disease proximities, and the quadratic terms capture possible adverse effects introduced by pairwise drug interactions.
Thus, we propose as problem Hamiltonian 
\begin{equation}  
\label{eq:Problem H}
H_Y = \sum_{i=1}^n \left(z(Y, X_i) + \beta\right) S_i - \gamma \sum_{i<j} s(X_i, X_j) S_i S_j.
\end{equation}  
The hyperparameter \( \beta \) acts as a bias towards combinations of a smaller  number of drugs (which are preferred in most scenarios), while \( \gamma \) balances the relative weight of pairwise interactions.  
In the language of Eq.~\eqref{eq:qubo_general}, this is equivalent to constructing a QUBO matrix with entries $Q_{ii} = z(Y, X_i) + \beta$ and $Q_{ij} = -\gamma\, s(X_i, X_j)$. Since a negative $z$-score reflects proximity to the disease module, it lowers $Q_{ii}$ favouring inclusion of the respective drug. For the quadratic terms, a positive $\gamma$ would ensure that separated drug pairs ($s(A,B) \geq 0$) contribute a negative weight to the cost function (lowering the energy of the final combination when both are co-selected) whereas overlapping pairs are penalised.  The idea is that for $\gamma, \beta > 0$, the ground state of $H_Y$ would precisely contain the configurations which best abide by the Complementary Exposure principle.

Furthermore, clinical feasibility and computational practicality often necessitate limiting the number of drugs in a combination. 
It is possible to enforce the selection of exactly \( k \) drugs  by appending a quadratic penalty term like
\begin{equation}  
H_{\text{constraint}} = \lambda \left(\sum_{i=1}^n S_i - k\right)^2.
\end{equation}  
A high value for \( \lambda \) ensures the ground state satisfies \( \sum_i S_i = k \). 
For instance, fixing \( k = 2 \) restricts solutions to dual-drug therapies, which dominate clinical practice due to their reduced risk of adverse interactions and more manageable testing \cite{Cheng2019}.
By introducing auxiliary variables, it is also possible to enforce multiple allowed combination sizes. In the special case where the two allowed sizes are consecutive integers, no ancilla are required and one can use the simpler term
\begin{equation}  
H_{\text{constraint}} = \lambda \left(\sum_{i=1}^n S_i - k_1\right)\left(\sum_{i=1}^n S_i - k_2\right),
\label{eq:dbl_constraint}
\end{equation}  
which penalizes all solutions except those with \( k_1 \) or \( k_2 \) drugs as long as \(|k_1-k_2|\leq1\).

The complete optimisation problem is then given by the full Hamiltonian $H_Y + H_{\text{constraint}}$, whose ground state encodes the optimal drug combination for disease $Y$ subject to the chosen cardinality constraints.

\section{Hyperparameter calibration and benchmarking}
\label{sec:calibration}
We now describe the procedure used throughout this work to identify promising drug combinations using the problem Hamiltonian formalism.
We subsequently apply this methodology to various diseases (see Section~\ref{sec:multiple_diseases}
for results).

In order to choose a suitable pair of hyperparameters $(\gamma, \beta)$ for the problem Hamiltonian \eqref{eq:Problem H} of a particular disease, we assembled a dataset consisting of validated drug combinations for various diseases.
This constructed benchmarking dataset was obtained by filtering known effective drug combinations to retain only those whose individual constituents have documented disease associations. Further details, including data sources and preprocessing steps, are discussed in Appendix \ref{sec:datasets}.  
The resulting dataset, denoted by $\mathcal{C}$, serves as a reference to evaluate the performance of different hyperparameter choices in prioritizing known effective drug combinations. 
From the approved drug combinations in $\mathcal{C}$, we build a candidate drug set $D$ by extracting unique drugs that appear in at least one combination in $\mathcal{C}$.
Comparing with the interactome data, we can determine the couplings $z(Y,X_i)$ and $s(X_i,X_j)$ for all $X_i, X_j\in D$ (here $Y$ refers to the chosen disease to be analysed).
Then, for any pair $(\gamma,\beta)$, we can evaluate the spectrum of the problem Hamiltonian.
As $D$ is relatively small, it is feasible to enumerate all $2^{|D|}$ configurations exactly, and rank them by their energy.

We evaluate the goodness of fit for a given pair $(\gamma,\beta)$ with an average precision (AP) metric. AP is particularly suitable here, as it measures the concentration of validated combinations among the lowest-energy states and handles imbalanced datasets effectively.
In more detail, let $c_i$ be the $i$-th configuration in the energy ranking (with $i=1$ corresponding to the lowest energy configuration), and define a binary label $y_i$ as
\begin{equation}
\label{eq:indicator}
y_i = 
\begin{cases}
1, & c_i\in \mathcal{C}\\
0, & \text{otherwise.}
\end{cases}
\end{equation}
We say that the precision at rank $k$ is 
\begin{equation}
\label{eq:precision}
P(k) = \frac{1}{k}\sum_{i=1}^{k} y_i,
\end{equation}
and define the average precision of the spectrum as
\begin{equation}
\mathrm{AP} = \frac{1}{|\mathcal{C}|}\sum_{k=1}^{2^{|D|}} P(k) y_k\,.
\end{equation}
By construction, $\mathrm{AP}\in[0,1]$, with larger values indicating better prioritization of validated combinations among the lowest energy solutions.

Performing a grid search over the hyperparameter space we can find the optimal values $(\gamma^*, \beta^*)$ for the chosen disease $Y$ that maximize this metric.
In our simulations, we selected a penalty term with $\lambda_{\text{penalty}}=10\cdot \max(|z|, |\gamma\cdot s|)$ and incorporated a constraint of the form of Eq.~\eqref{eq:dbl_constraint} with $k_1=2$ and $k_2=3$.
This strongly favours combinations of size $2$ and $3$, which make up the bulk of the validated combinations dataset for all diseases in $\mathcal{C}$.

Having selected the hyperparameters, we seek to apply our model to a larger candidate drug set $\mathcal{D}$, containing drugs that have not been validated in any combined therapy.
This enlarged set is constructed by padding the original pool with randomly selected additional drugs, while ensuring that no strongly overlapping drugs lacking validated combination information are included.
The size of this resulting set $|\mathcal{D}|$ needs to be chosen to match the practical embedding and sampling limits of the available hardware, ensuring that the optimisation remains technically viable.
By enlarging $\mathcal{D}$, we are populating the low-energy landscape to admit both known effective combinations and novel candidates whose network proximities to other drugs and the disease module resemble those of validated therapies.
Thus, applying the learned hyperparameters $(\gamma^*,\beta^*)$ to the QUBO constructed for the expanded candidate set $\mathcal{D}$ allows us to identify potential new drug combinations, including those involving previously untested drugs. We demonstrate how we can extract such predictions in the next section.

\section{Drug Combination Prediction}

In an ideal adiabatic process, the annealer is expected to return only the ground state configuration. However, in practice, hardware limitations can prevent the adiabatic evolution condition (\ref{eq:time_scaling}) from being satisfied. Together with uncontrollable noise, this causes the device to produce a distribution spanning many low-energy states, rather than a single ground state.

This aligns well with our goal of extracting multiple candidate combinations. We can leverage the non-idealities of the annealing process to explore a variety of low-energy configurations and thereby obtain many near-optimal drug combinations, which can be ranked and prioritised for experimental validation.
To reproduce this behaviour on the previously mentioned enlarged QUBO ($\mathcal{D}$), we employ an SQA sampler, D-Wave's \texttt{PathIntegralAnnealingSampler} \cite{dwavePathIntegralAnnealingSampler}, to reproduce the marginal distribution over low-energy states produced by a physical annealer.
Rather than the sampler's default geometric schedule, we define a custom exponential annealing schedule for the problem and driver field strengths, motivated by the idea that custom annealing schedules can enhance the diversity of sampled low-energy solutions \cite{PhysRevE.108.065303}. The problem-field schedule $\{H_p(t_k)\}_{k=0}^{N}$ is constructed as
\begin{equation}
    H_p(t_k) = \beta_{\max}\,
    \frac{e^{\alpha\, k/N} - 1}{e^{\alpha} - 1}\,,
    \qquad k = 0, 1, \ldots, N\,,
\end{equation}
with $\beta_{\max} = 3.0$, $\alpha = 1.5$, and $N = 10{,}001$ steps, while the driver-field schedule is set to $H_d(t_k) = H_p(t_{N-k})$, realising an exponentially accelerating transition. Two sweeps are performed per schedule step (\texttt{num\_sweeps\_per\_beta} $= 2$), yielding a total of $2N$ single-spin-flip sweeps per read. This schedule is heuristically designed to keep the system near low-energy states during the slow early phase while allowing rapid thermalisation into the problem's ground-state during the fast late phase. As a result, a considerable number of runs escape the ground state of the problem Hamiltonian and sample a variety of configurations with similar (low) energy, which is precisely the regime where near-optimal drug combinations lie.

Following this scheme, the sampler is run $R$ independent times. Each possible combination $c$ will have been observed $n_{c}$ times with corresponding frequency $f_{c}\;=\;n_{c}/R$. We then require a sufficiently large number of runs $R$ so that $f_c>0$ for the combinations in the region of interest. With the majority of the probability mass concentrated on low-energy states, we can filter the lowest energy (generally highest frequency) configurations, which were not part of the validated combinations dataset, to set as our model's predictions for potential novel drug combinations and rank them accordingly. 

This produces a ranked list of plausible novel combinations whose network-based profiles resemble those of validated therapies.
Taken together, this approach provides a practical mechanism for extracting multiple candidate combinations from the quantum annealing process, effectively leveraging the hardware's characteristics to our advantage.

\section{Results}
\label{sec:multiple_diseases}
We apply the Hamiltonian framework and hyperparameter calibration pipeline to the four diseases with the most documented validated combinations in the benchmark dataset $\mathcal{C}$. The results below follow the methodology described previously.

Firstly, for each disease, we construct a set of candidates $D$ from the unique drugs that make up its validated drug combinations. For the diseases Diabetes Mellitus and Rheumatoid Arthritis, these candidate sets consist of $10$ unique drugs, while for Asthma and Brain Neoplasms $9$ such drugs are considered. 

Using the interactome data, for each disease $Y$ and drug pair $X_i, X_j\in D$, we compute the network measures $z(Y,X_i)$ and $s(X_i,X_j)$, which are then used to construct test QUBOs for specific hyperparameter pairs.  
\begin{figure*}[htbp]
    \centering
    \includegraphics[width=\textwidth]{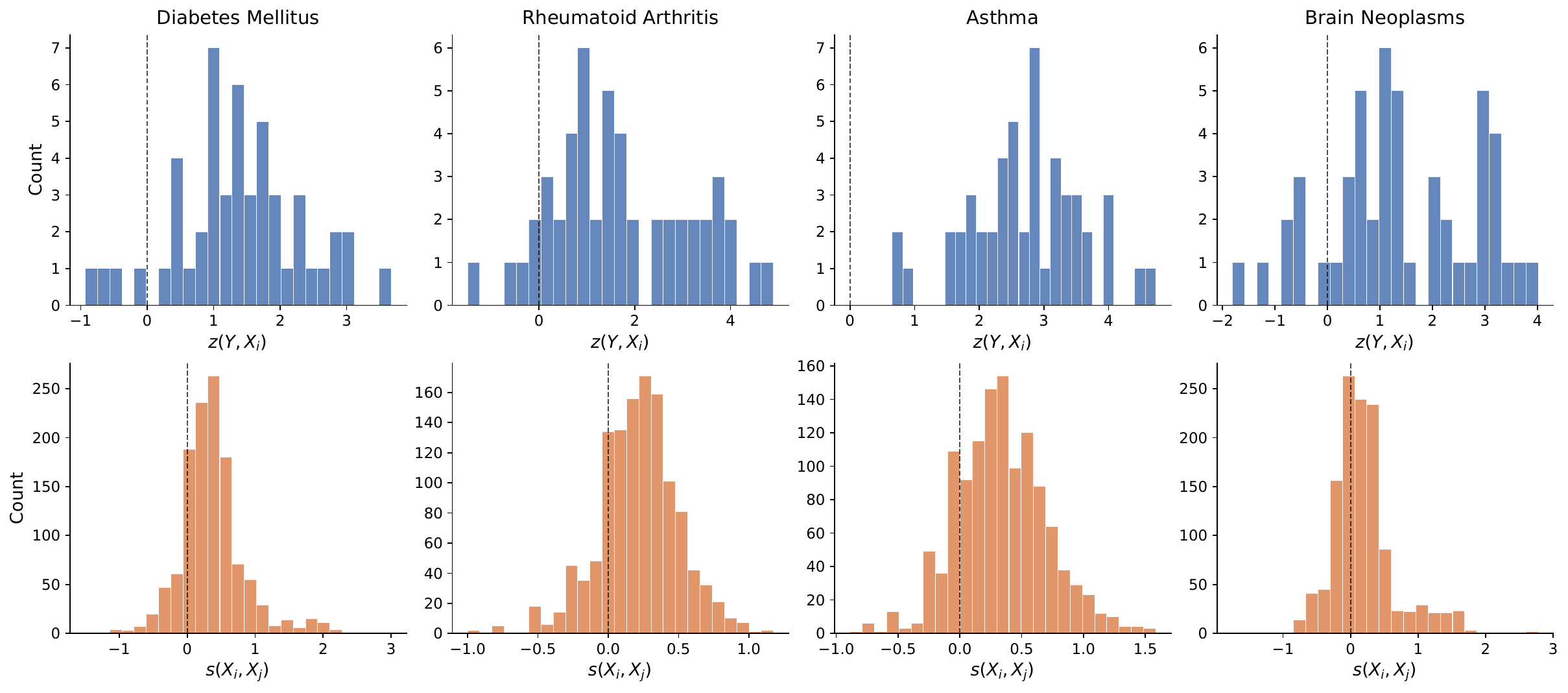}
    \caption{\justifying \textbf{Empirical distributions of network proximity metrics.} 
    Distributions of the drug-disease proximity $z$-score, $z(Y, X_i)$ (top row), for a subset of 50 candidate drugs evaluated against four distinct diseases (Diabetes Mellitus, Rheumatoid Arthritis, Asthma, and Brain Neoplasms) and the corresponding distributions of the drug-drug network separation, $s(X_i, X_j)$, for all pairs within the 50 drug sets.}
    \label{fig:zs_distributions}
\end{figure*}
As illustrated in Figure~\ref{fig:zs_distributions}, an analysis of the empirical distributions of these metrics reveals that the vast majority of candidate drugs exhibit $z > 0$, inherently penalizing their selection via positive linear QUBO terms. While drugs from the original validated sets typically populate the negative left tail ($z < 0$), Asthma presents a notable exception where all of its candidate drugs, including drugs from verified combinations, maintain $z > 0$. This illustrates the difficulty of capturing topological proximity for certain diseases and increasing the reliance on precise parameter tuning and hence dataset quality. 

Conversely, the $s$-score distributions highlight the sparsity of target overlap. Because few drug pairs have $s < 0$, most off-diagonal QUBO entries evaluate to negative values, naturally rewarding the co-selection of separated drug pairs in line with the Complementary Exposure principle. 
Thus, the interplay between linear selection penalties and quadratic pairing rewards makes a cardinality constraint essential within the QUBO formulation for some diseases. Without strictly bounding the number of selected nodes, the objective function risks collapsing into trivial ground states, such as selecting all disjoint candidate drugs or none at all. Enforcing a strict combination size prevents this runaway selection and enables the inclusion of combinations that are not rigidly bound by the hard sign cutoffs ($z < 0$ and $s < 0$) of the individual metrics.

By sweeping the hyperparameters $(\gamma,\beta)$ and computing the AP for each pair, we are able to determine the optimal parameters $(\gamma^*,\beta^*)$ that compensate for these unique network topologies and maximize the AP metric.

The calibrated hyperparameter pairs $(\gamma^*,\beta^*)$ and their corresponding AP values for each disease are given in Table~\ref{tab:hyperparams}.

\begin{figure*}[htbp]
  \centering

  \begin{subfigure}[b]{0.48\textwidth}
    \refstepcounter{subfigure}\label{fig:subapa}%
    \begin{tikzpicture}
      \node[inner sep=0pt] (img) {\includegraphics[width=\linewidth]{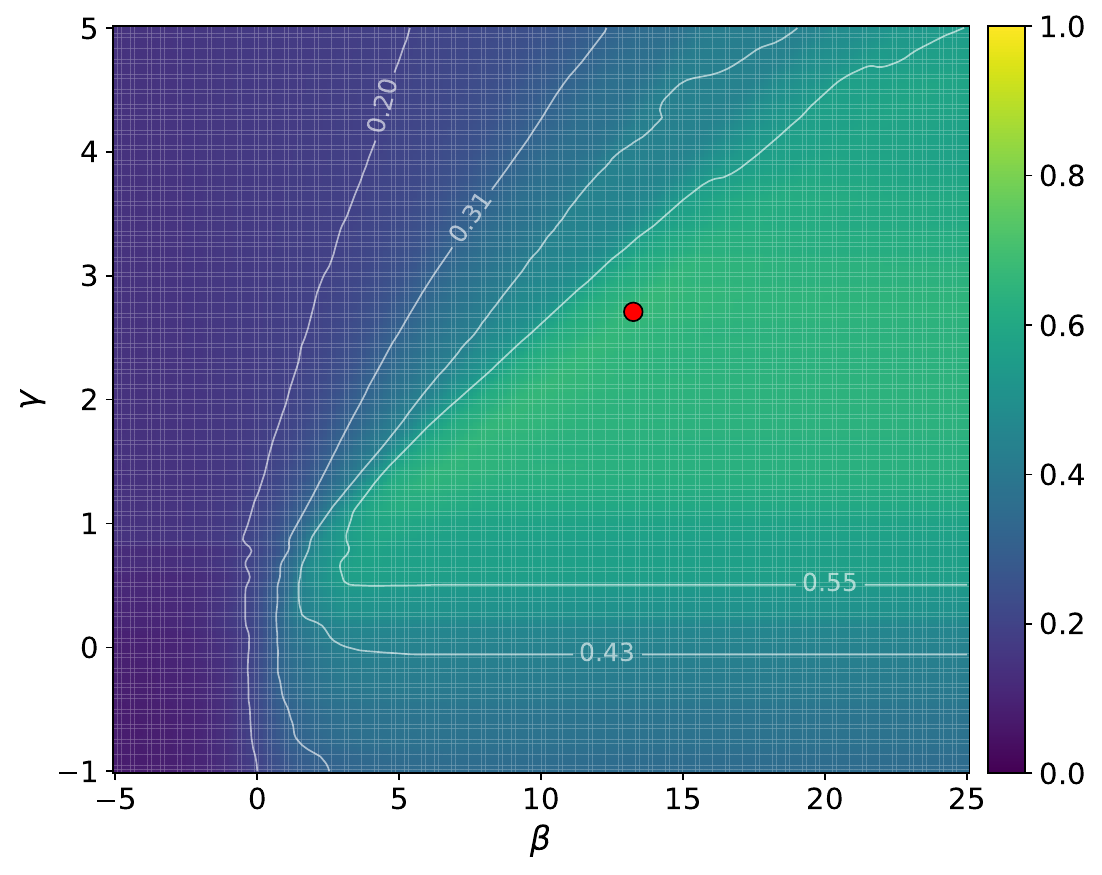}};
      \node[anchor=north west, yshift=10pt,fill=white,inner sep=2pt,font=\bfseries] at (img.north west) {(\thesubfigure)};
    \end{tikzpicture}
  \end{subfigure}\hfill
  \begin{subfigure}[b]{0.48\textwidth}
    \refstepcounter{subfigure}\label{fig:subapb}%
    \begin{tikzpicture}
      \node[inner sep=0pt] (img) {\includegraphics[width=\linewidth]{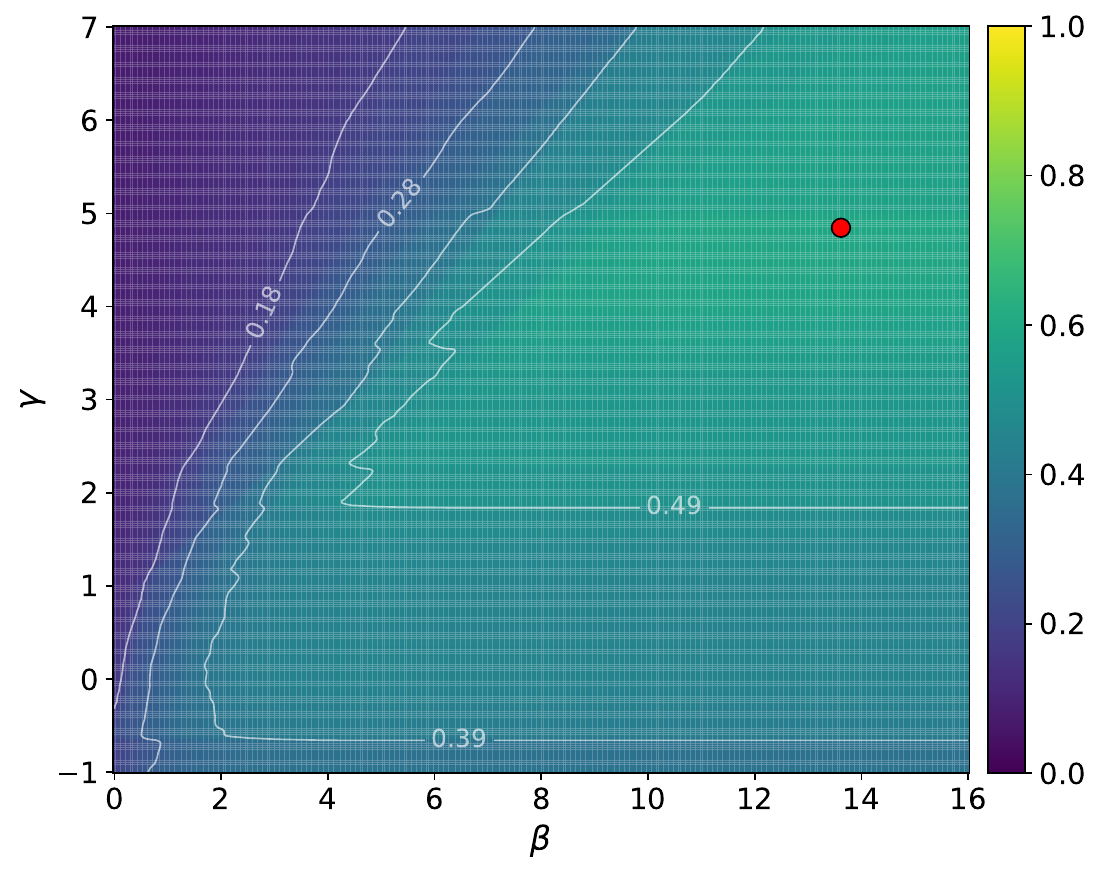}};
      \node[anchor=north west, yshift=10pt,fill=white,inner sep=2pt,font=\bfseries] at (img.north west) {(\thesubfigure)};
    \end{tikzpicture}
  \end{subfigure}

  \vspace{2mm}

  \begin{subfigure}[b]{0.48\textwidth}
    \refstepcounter{subfigure}\label{fig:subapc}%
    \begin{tikzpicture}
      \node[inner sep=0pt] (img) {\includegraphics[width=\linewidth]{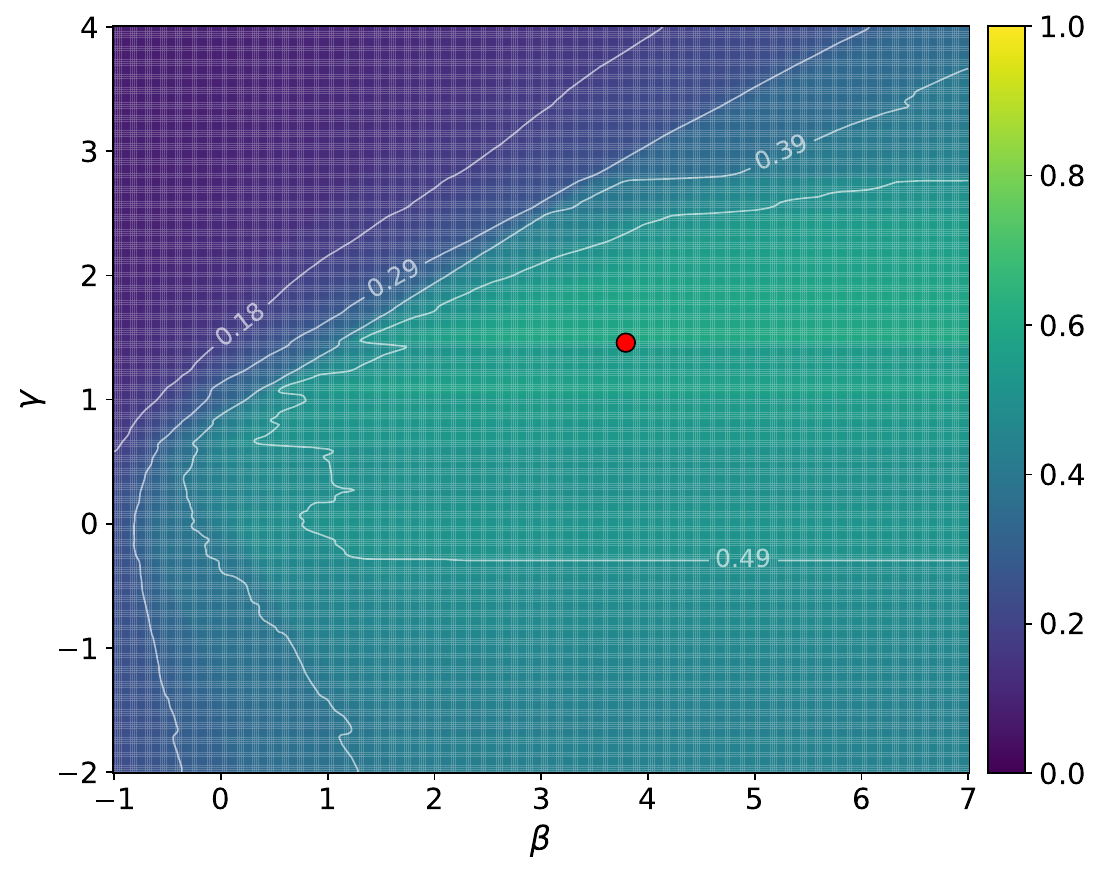}};
      \node[anchor=north west, yshift=10pt,fill=white,inner sep=2pt,font=\bfseries] at (img.north west) {(\thesubfigure)};
    \end{tikzpicture}
  \end{subfigure}\hfill
  \begin{subfigure}[b]{0.48\textwidth}
    \refstepcounter{subfigure}\label{fig:subapd}%
    \begin{tikzpicture}
      \node[inner sep=0pt] (img) {\includegraphics[width=\linewidth]{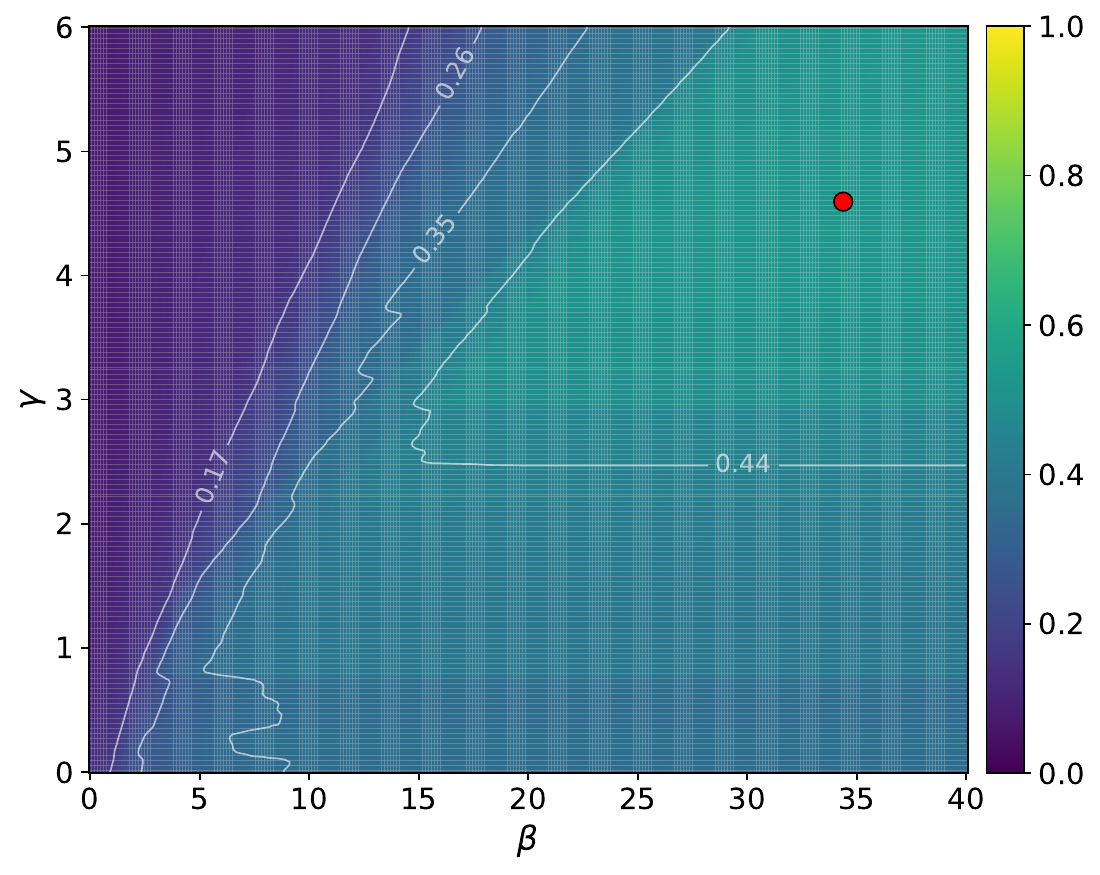}};
      \node[anchor=north west, yshift=10pt,fill=white,inner sep=2pt,font=\bfseries] at (img.north west) {(\thesubfigure)};
    \end{tikzpicture}
  \end{subfigure}

  \caption{\justifying \textbf{Average Precision (AP) Landscape.} Each subplot displays the AP values over a grid of hyperparameters $(\gamma, \beta)$ for the different diseases: \textbf{(a)} Diabetes Mellitus, \textbf{(b)} Rheumatoid Arthritis, \textbf{(c)} Asthma, and \textbf{(d)} Brain Neoplasms. The red dot in each plot indicates the optimal hyperparameter pair $(\gamma^*, \beta^*)$ that maximizes the AP metric, reflecting the best prioritization of validated drug combinations among low-energy configurations.}
\label{fig:ap_landscape}
\end{figure*}

\begin{table}[t]
\centering
\caption{Calibrated hyperparameters $(\gamma^*,\beta^*)$ and corresponding Average Precision (AP) for each disease.}
\label{tab:hyperparams}
\setlength{\tabcolsep}{8pt}      
\renewcommand{\arraystretch}{1.25}
\begin{tabular}{lccc}
\toprule
Disease & $\gamma^*$ & $\beta^*$ & $\mathrm{AP}$ \\
\midrule
Diabetes Mellitus    & $2.74$ & $13.39$ & $0.670$ \\
Rheumatoid Arthritis & $5.18$ & $14.47$ & $0.594$ \\
Asthma               & $1.46$ & $3.79$  & $0.600$ \\
Brain Neoplasms      & $4.60$ & $34.38$ & $0.528$ \\
\bottomrule
\end{tabular}
\end{table}
The AP landscapes in the neighborhoods of these optimal points are displayed in Figure ~\ref{fig:ap_landscape}, showing a visible bias towards the first quadrant of the $(\gamma,\beta)$ plane. This is consistent with the Complementary Exposure principle and further validates our Hamiltonian formulation. 
Indeed, positive $\gamma$ rewards target separation while positive $\beta$ penalises large combinations. 
The AP landscapes in Figure~\ref{fig:ap_landscape} reinforce this, as AP degrades monotonically away from the first quadrant and effectively vanishes for $\beta \leq 0$ across all four diseases, confirming that meaningful prioritisation requires both parameters to encode the intended biological logic.

\begin{figure*}[htbp]
  \centering

  \begin{subfigure}[b]{0.48\textwidth}
    \refstepcounter{subfigure}\label{fig:subspa}%
    \begin{tikzpicture}
      \node[inner sep=0pt] (img) {\includegraphics[width=\linewidth]{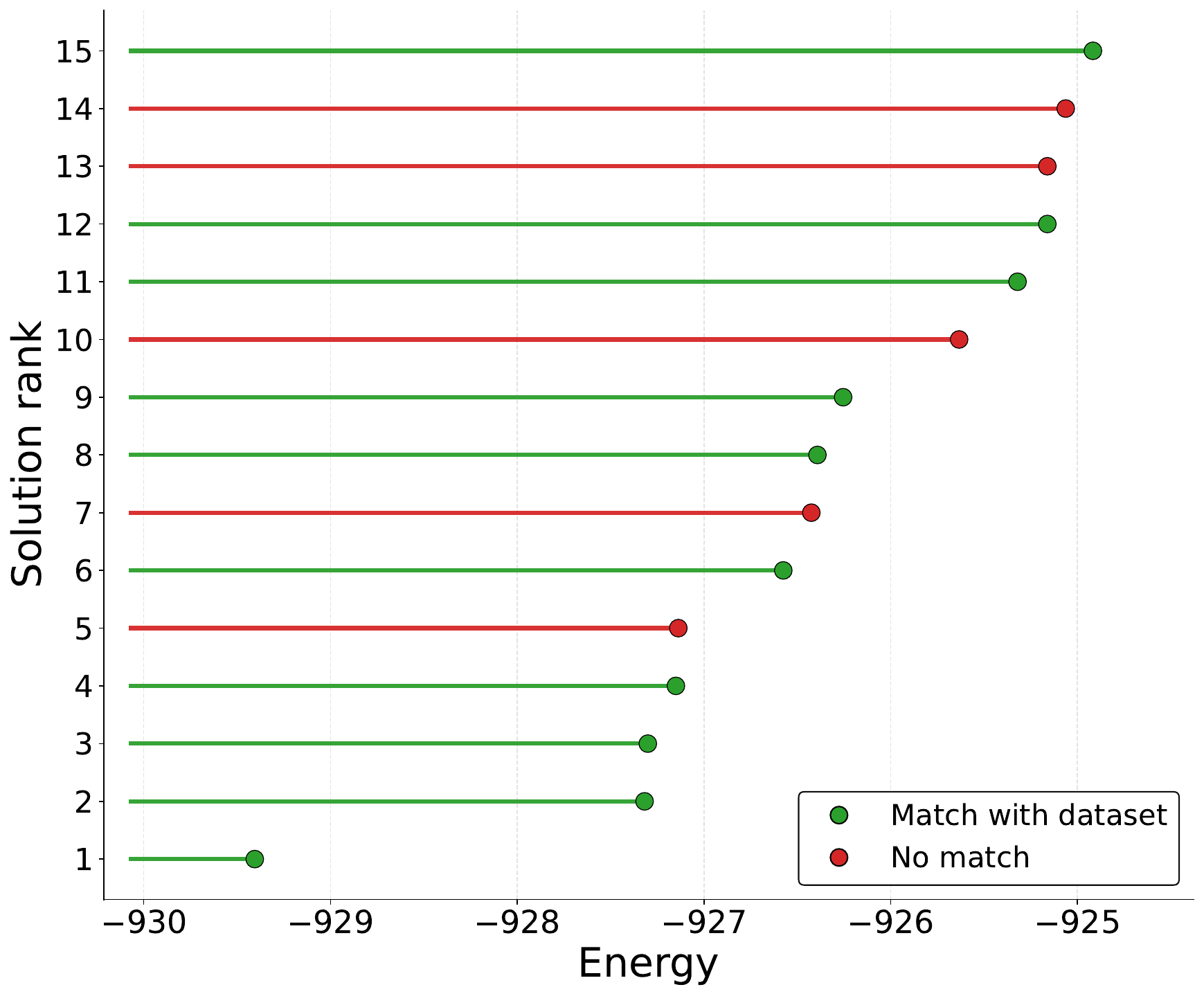}};
      \node[anchor=north west, yshift=10pt,fill=white,inner sep=2pt,font=\bfseries] at (img.north west) {(\thesubfigure)};
    \end{tikzpicture}
  \end{subfigure}\hfill
  \begin{subfigure}[b]{0.48\textwidth}
    \refstepcounter{subfigure}\label{fig:subspb}%
    \begin{tikzpicture}
      \node[inner sep=0pt] (img) {\includegraphics[width=\linewidth]{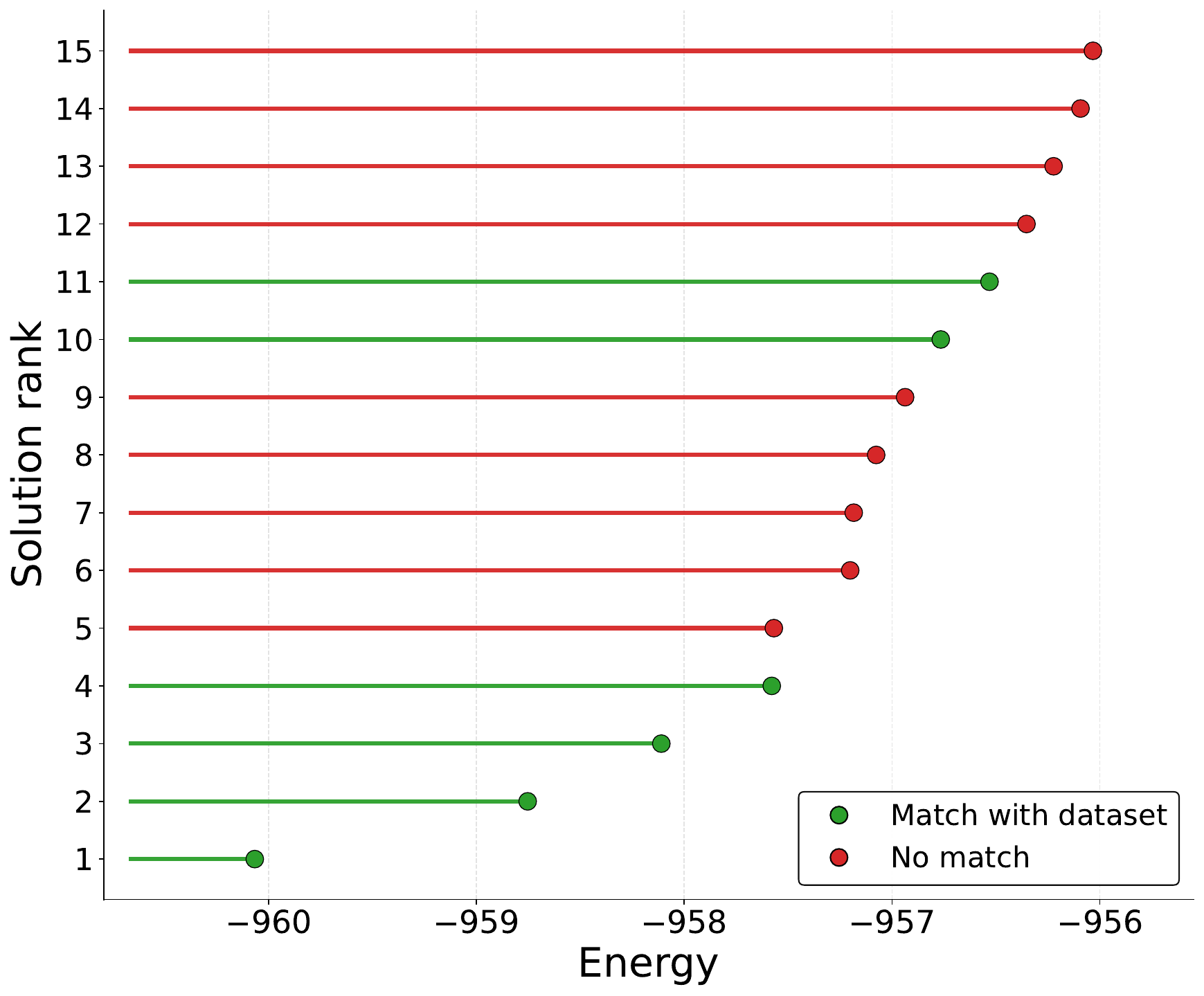}};
      \node[anchor=north west, yshift=10pt,fill=white,inner sep=2pt,font=\bfseries] at (img.north west) {(\thesubfigure)};
    \end{tikzpicture}
  \end{subfigure}

  \vspace{2mm}

  \begin{subfigure}[b]{0.48\textwidth}
    \refstepcounter{subfigure}\label{fig:subspc}%
    \begin{tikzpicture}
      \node[inner sep=0pt] (img) {\includegraphics[width=\linewidth]{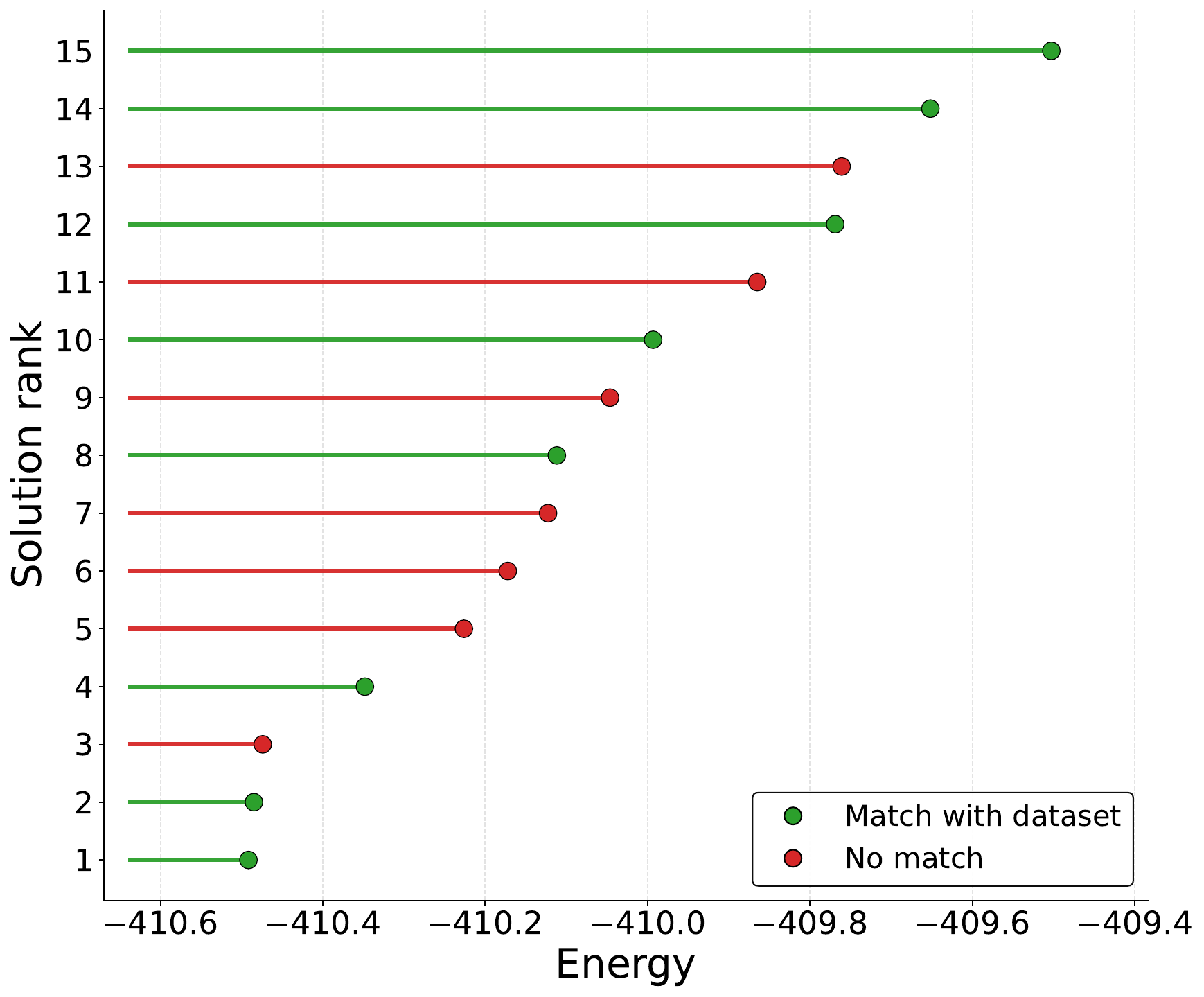}};
      \node[anchor=north west, yshift=10pt,fill=white,inner sep=2pt,font=\bfseries] at (img.north west) {(\thesubfigure)};
    \end{tikzpicture}
  \end{subfigure}\hfill
  \begin{subfigure}[b]{0.48\textwidth}
    \refstepcounter{subfigure}\label{fig:subspd}%
    \begin{tikzpicture}
      \node[inner sep=0pt] (img) {\includegraphics[width=\linewidth]{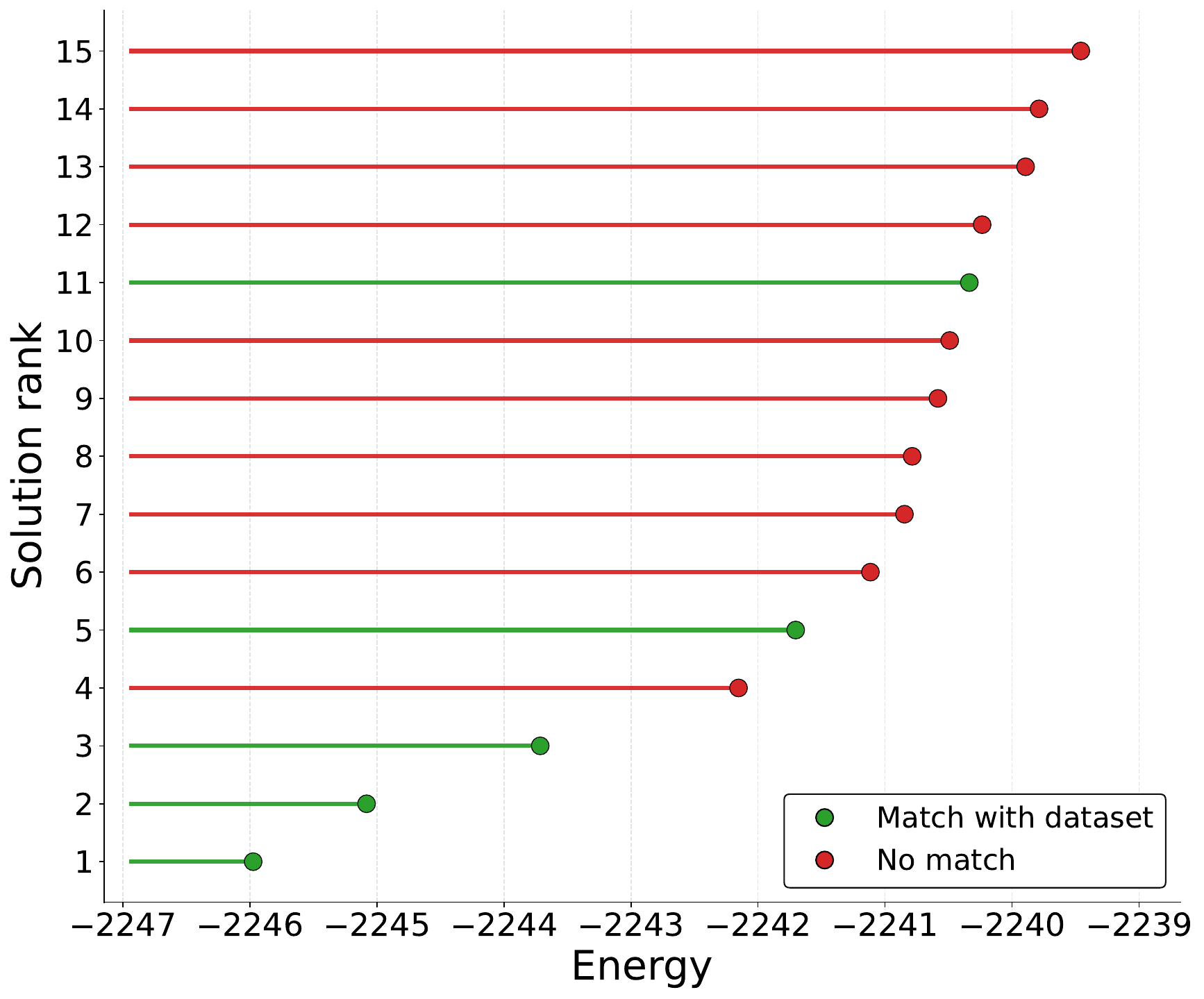}};
      \node[anchor=north west, yshift=10pt,fill=white,inner sep=2pt,font=\bfseries] at (img.north west) {(\thesubfigure)};
    \end{tikzpicture}
  \end{subfigure}

  \caption{\justifying \textbf{Energy spectrum at the optimal hyperparameters} Each subplot displays the top 15 lowest-energy configurations of the QUBO constructed from the initial candidate drug set at the disease-specific optimal hyperparameters $(\gamma^*, \beta^*)$. Solutions are ordered by increasing QUBO energy (lowest at the bottom) and are colour-coded to indicate whether the configuration matches a validated combination in the reference set $\mathcal{C}$ or not. The diseases represented are: \textbf{(a)} Diabetes Mellitus, \textbf{(b)} Rheumatoid Arthritis, \textbf{(c)} Asthma, and \textbf{(d)} Brain Neoplasms.}
\label{fig:spectrum}
\end{figure*}

To better visualize the prioritization of the validated combinations at the calibrated point, we inspect the bottom of the QUBO energy spectrum. Figure \ref{fig:spectrum} shows how the validated combinations populate the low-energy region. The most pronounced structure is found for Diabetes Mellitus, which is also the disease with the largest number of documented combinations in $\mathcal{C}$. The observed tight clustering of matched combinations corresponds to a high AP and suggests that the network-proximity and separation metrics capture salient features of efficacious combinations. Conversely, higher dispersion (and lower AP) is indicative of either sparse benchmark data or disease biology that is less amenable to pairwise topological description. Namely, in the case of Brain Neoplasms, which holds the lowest AP, only one of the combinations present in $\mathcal{C}$ contains three drugs (the remainder being pairwise combinations). This imbalance in combination sizes is reflected in a considerably higher $\beta^*$ value, which in turn greatly expands the energy range of the respective QUBO. This increase in scale usually tightens the spectral gap and can hinder optimisation when considering high numbers of drugs. These observations reinforce the need for a well-curated benchmark dataset and directly inform how confidently calibrated parameters can be used to search larger candidate spaces, since concentrated spectra imply more confidently that low-energy configurations are meaningful targets for drug discovery, while interleaved spectra could be less trustworthy, requiring more conservative interpretations and motivating additional data collection.

\begin{figure*}[htbp] 
  \centering

  \begin{subfigure}[b]{0.48\textwidth}
    \refstepcounter{subfigure}\label{fig:subpra}%
    \begin{tikzpicture}
      \node[inner sep=0pt] (img) {\includegraphics[width=\linewidth]{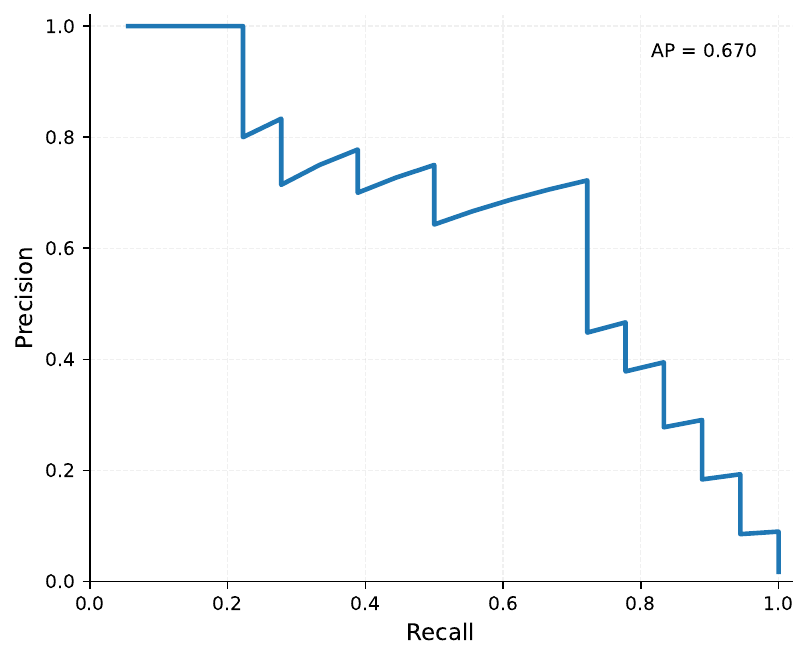}};
      \node[anchor=north west, yshift=10pt,fill=white,inner sep=2pt,font=\bfseries] at (img.north west) {(\thesubfigure)};
    \end{tikzpicture}
  \end{subfigure}\hfill
  \begin{subfigure}[b]{0.48\textwidth}
    \refstepcounter{subfigure}\label{fig:subprb}%
    \begin{tikzpicture}
      \node[inner sep=0pt] (img) {\includegraphics[width=\linewidth]{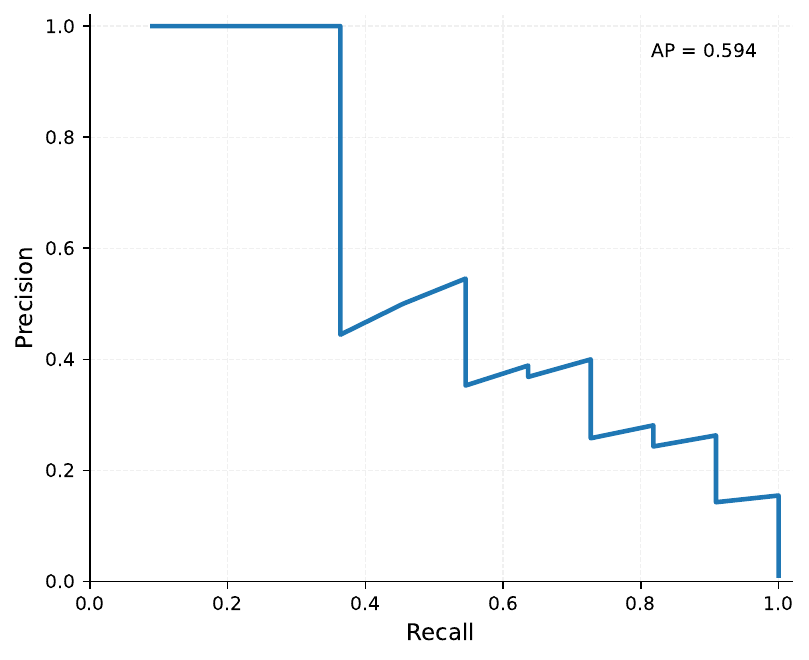}};
      \node[anchor=north west, yshift=10pt,fill=white,inner sep=2pt,font=\bfseries] at (img.north west) {(\thesubfigure)};
    \end{tikzpicture}
  \end{subfigure}

  \vspace{2mm}

  \begin{subfigure}[b]{0.48\textwidth}
    \refstepcounter{subfigure}\label{fig:subprc}%
    \begin{tikzpicture}
      \node[inner sep=0pt] (img) {\includegraphics[width=\linewidth]{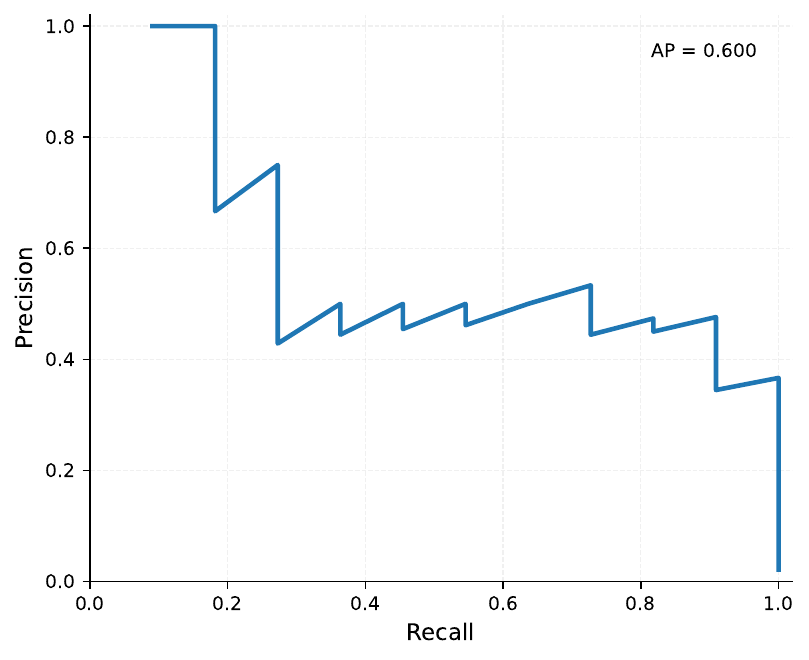}};
      \node[anchor=north west, yshift=10pt,fill=white,inner sep=2pt,font=\bfseries] at (img.north west) {(\thesubfigure)};
    \end{tikzpicture}
  \end{subfigure}\hfill
  \begin{subfigure}[b]{0.48\textwidth}
    \refstepcounter{subfigure}\label{fig:subprd}%
    \begin{tikzpicture}
      \node[inner sep=0pt] (img) {\includegraphics[width=\linewidth]{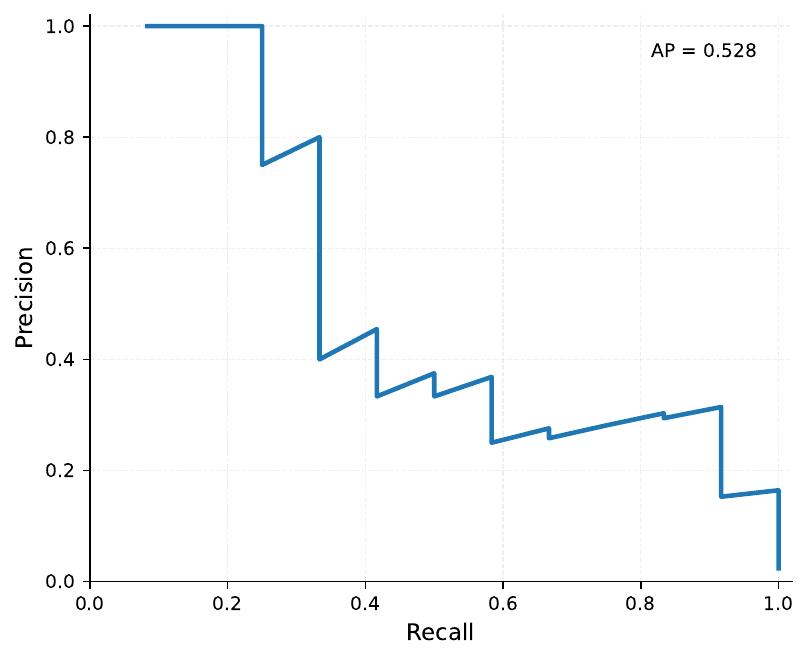}};
      \node[anchor=north west, yshift=10pt,fill=white,inner sep=2pt,font=\bfseries] at (img.north west) {(\thesubfigure)};
    \end{tikzpicture}
  \end{subfigure}

  \caption{\justifying \textbf{Precision-Recall Curve at the optimal hyperparameters.} Each subplot displays the Precision-Recall curve obtained by sweeping the ranked solutions of the QUBO constructed from the initial candidate drug set at the disease-specific optimal hyperparameters $(\gamma^*, \beta^*)$. The area under each curve corresponds to the Average Precision (AP) and quantifies the prioritization of validated drug combinations among the low-energy configurations. The diseases represented are: \textbf{(a)} Diabetes Mellitus, \textbf{(b)} Rheumatoid Arthritis, \textbf{(c)} Asthma, and \textbf{(d)} Brain Neoplasms.}
\label{fig:prcurve}
\end{figure*}

A visual depiction of how the representativeness of matched combinations degrades as we move up in energy and rank is provided by the Precision-Recall curves in Figure~\ref{fig:prcurve}. 
Recall at rank $k$ is defined as the fraction of all validated combinations retrieved within the top $k$ configurations,
\begin{equation}
R(k) = \frac{1}{|\mathcal{C}|}\sum_{i=1}^{k} y_i\,,
\end{equation}
where $y_i$ is the indicator defined in Eq.~\eqref{eq:indicator}. 
The Precision-Recall curve then plots $P(k)$ (Eq.~\eqref{eq:precision}) against $R(k)$ as $k$ is swept from $1$ to the total number of configurations, with the area under the curve corresponding to the AP.
The number of ground-truth positives ranges from $11$--$18$, a small number compared to the total number of configurations, which explains the jaggedness of the curves. 

For discovery, the calibrated hyperparameters \((\gamma^*,\beta^*)\) are then applied to example enlarged candidate sets $\mathcal{D}$ of size $|\mathcal{D}|=50$ (the previous ones plus $40$ or $41$ new random drugs). We run the SQA Sampler for $1024$ independent runs, a quantity commonly used in quantum annealing experiments. 
In addition to the Average Precision and Precision at Rank $k$ ($P(k)$) established in Section~\ref{sec:calibration}, we introduce two complementary metrics to assess prediction quality on the enlarged candidate sets. 
While AP and $P(k)$ are well suited for evaluating the ranking produced by the deterministic energy spectrum on the small calibration set, in the enlarged setting the candidate space is now dominated by a vast majority of unvalidated configurations, making it useful to quantify how much better than random the ranking performs.
Furthermore, the predictions are no longer extracted from a deterministic ranking but from the empirical frequencies of a stochastic SQA sampler, whose probability mass distribution over low-energy states is itself informative.
First, we define the \emph{enrichment ratio} at rank $k$.
Let $\mathcal{K} = \{k_1, k_2\}$ denote the set of allowed combination sizes imposed by the cardinality constraint (Eq.~\ref{eq:dbl_constraint}), and let
\begin{equation}
N_{\text{total}} = \sum_{k \in \mathcal{K}} \binom{|D|}{k}
\end{equation}
be the total number of candidate configurations admitted by this constraint. 
The base rate of validated combinations is then $r_0 = |\mathcal{C}| / N_{\text{total}}$, and the enrichment ratio at rank~$k$ is defined as
\begin{equation}
E(k) = \frac{P(k)}{r_0}\,.
\end{equation}
An enrichment ratio much larger than unity indicates that the algorithm concentrates validated combinations at the top of the ranking far beyond what random ordering would achieve. 
Second, for the SQA sampling results we report the \emph{cumulative match frequency} $f_{\text{match}}$, defined as the total fraction of the $R = 1024$ independent SQA runs whose output corresponds to a validated combination among the top~$k$ lowest-energy configurations. 
This metric quantifies the extent to which the sampler's probability mass concentrates on known effective therapies.

\begin{figure*}[htbp]
  \centering

  \begin{subfigure}[b]{0.48\textwidth}
    \refstepcounter{subfigure}\label{fig:subsqaa}%
    \begin{tikzpicture}
      \node[inner sep=0pt] (img) {\includegraphics[width=\linewidth]{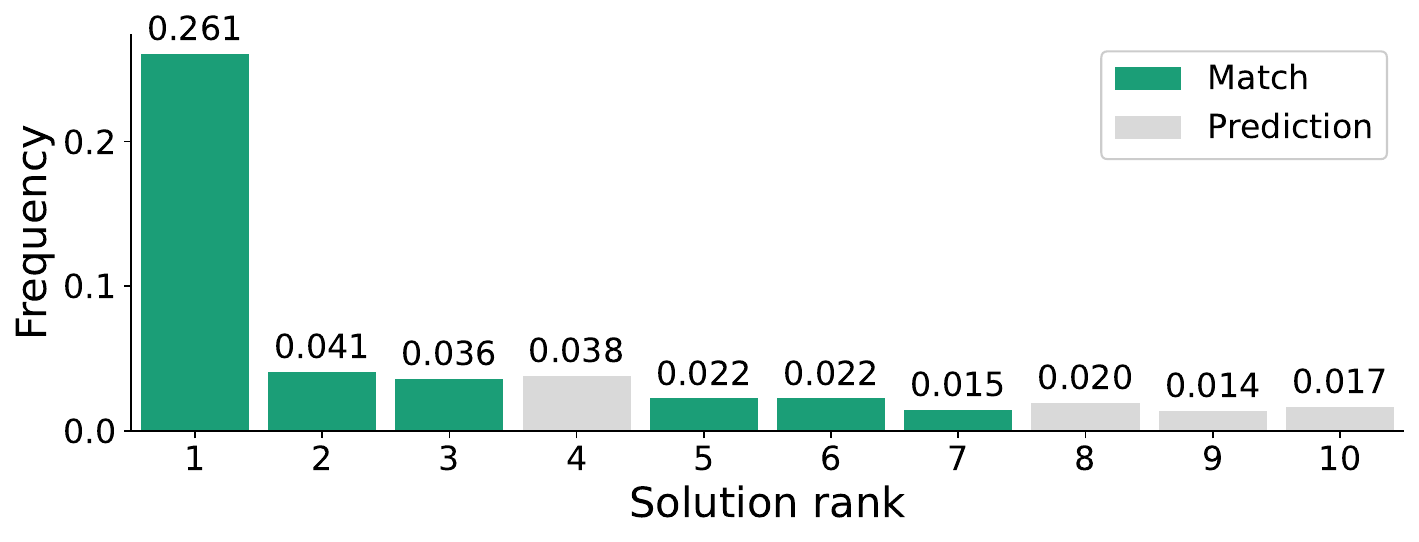}};
      \node[anchor=north west, yshift=10pt,fill=white,inner sep=2pt,font=\bfseries] at (img.north west) {(\thesubfigure)};
    \end{tikzpicture}
  \end{subfigure}\hfill
  \begin{subfigure}[b]{0.48\textwidth}
    \refstepcounter{subfigure}\label{fig:subsqab}%
    \begin{tikzpicture}
      \node[inner sep=0pt] (img) {\includegraphics[width=\linewidth]{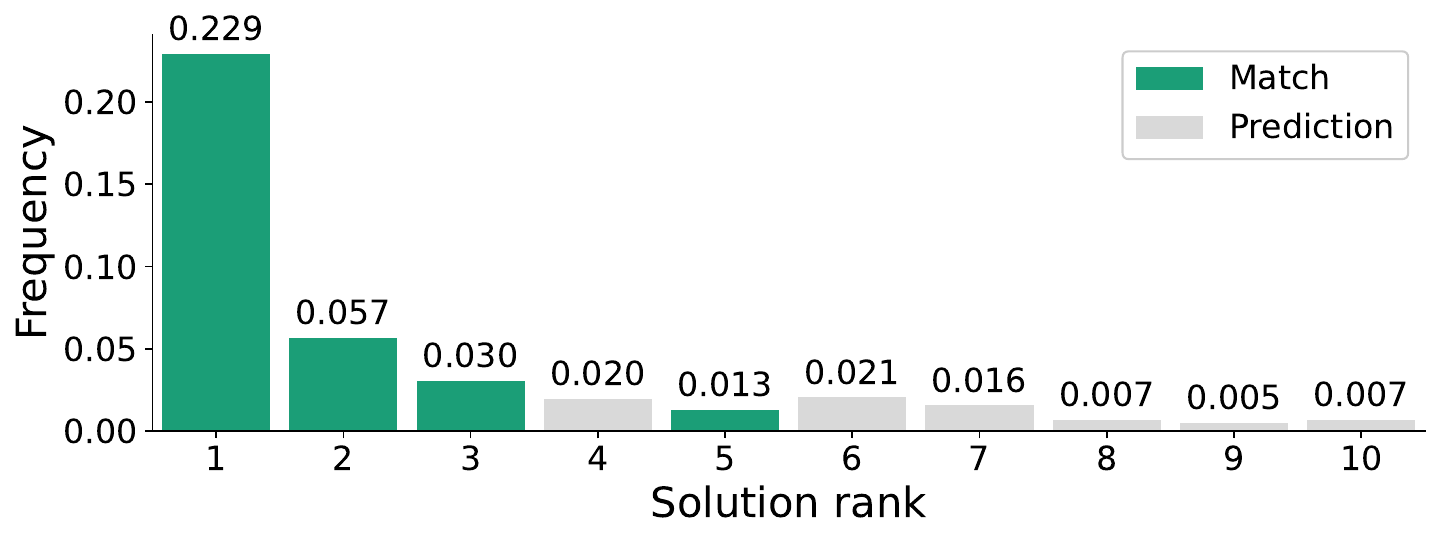}};
      \node[anchor=north west, yshift=10pt,fill=white,inner sep=2pt,font=\bfseries] at (img.north west) {(\thesubfigure)};
    \end{tikzpicture}
  \end{subfigure}

  \vspace{2mm}

  \begin{subfigure}[b]{0.48\textwidth}
    \refstepcounter{subfigure}\label{fig:subsqac}%
    \begin{tikzpicture}
      \node[inner sep=0pt] (img) {\includegraphics[width=\linewidth]{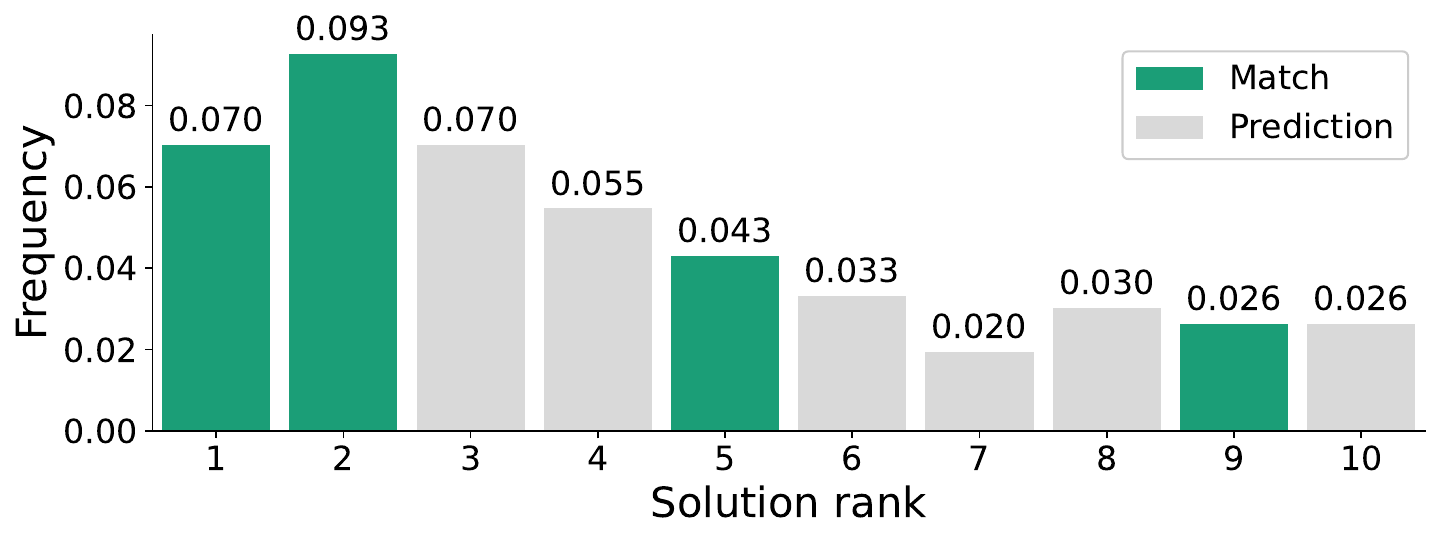}};
      \node[anchor=north west, yshift=10pt,fill=white,inner sep=2pt,font=\bfseries] at (img.north west) {(\thesubfigure)};
    \end{tikzpicture}
  \end{subfigure}\hfill
  \begin{subfigure}[b]{0.48\textwidth}
    \refstepcounter{subfigure}\label{fig:subsqad}%
    \begin{tikzpicture}
      \node[inner sep=0pt] (img) {\includegraphics[width=\linewidth]{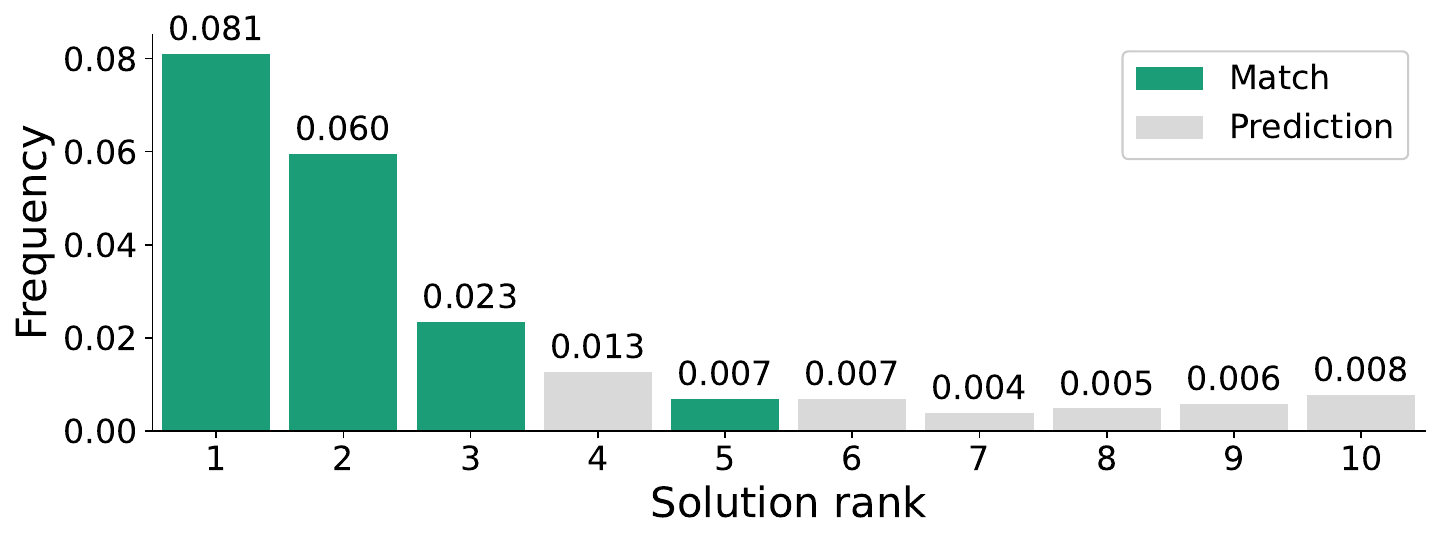}};
      \node[anchor=north west, yshift=10pt,fill=white,inner sep=2pt,font=\bfseries] at (img.north west) {(\thesubfigure)};
    \end{tikzpicture}
  \end{subfigure}

  \caption{\justifying \textbf{Sampled Simulated Quantum Annealing.} Each subplot displays the empirical frequency distribution from $1024$ runs over the $10$ lowest QUBO energies obtained by running the SQA sampler on the enlarged candidate drug set $\mathcal{D}$ at the disease-specific optimal hyperparameters $(\gamma^*, \beta^*)$. Green markers indicate combinations found in $\mathcal{C}$ while the remaining configurations are taken as predicted candidate synergistic combinations. The diseases represented are: \textbf{(a)} Diabetes Mellitus, \textbf{(b)} Rheumatoid Arthritis, \textbf{(c)} Asthma, and \textbf{(d)} Brain Neoplasms.}
\label{fig:SQA}
\end{figure*}

Table~\ref{tab:prediction_metrics} summarises the prediction quality on the enlarged candidate sets ($|D| = 50$) while Figure~\ref{fig:SQA} showcases the marginal distributions obtained from the SQA runs.
We observe that the produced empirical frequencies are spread across many low-energy states, some of which include previously untested drugs. 
In fact, for all four diseases, the candidate space comprises $N_{\text{total}} = 20{,}825$ configurations (on the order of $10^{15}$ possible drug combinations disregarding cardinality constraints), of which only $|\mathcal{C}|$ are validated, yielding base rates below $0.1\%$. 
Despite this extreme class imbalance, the lowest-energy configurations are densely populated by validated combinations. For instance, the precision at rank~5 reaches $P(5) = 0.80$ for Diabetes Mellitus, Rheumatoid Arthritis, and Brain Neoplasms, corresponding to enrichment ratios on the order of $10^3$. 
At rank $10$, the precision remains between $0.40$ and $0.60$ across all diseases, confirming that the calibrated Hamiltonian ranks validated combinations far above random expectation.

The SQA sampling results (Figure~\ref{fig:SQA}) corroborate this finding.
For Diabetes Mellitus, the cumulative frequency of validated combinations among the 10 lowest-energy sampled configurations is $f_{\text{match}} = 0.397$, meaning that approximately 40\% of all 1024 independent SQA runs returned a validated drug combination within a search space where such combinations constitute only 0.08\% of all candidates. 
Comparable concentration is observed for Rheumatoid Arthritis ($f_{\text{match}} = 0.329$) and, to a lesser extent, for Brain Neoplasms ($f_{\text{match}} = 0.171$) and Asthma ($f_{\text{match}} = 0.232$).
This suggests the calibrated hyperparameters generalize to larger candidate sets and points to possible therapeutic relevance of similar energy combinations.

\begin{table}[t]
\centering
\small
\caption{\justifying \textbf{Prediction quality on the enlarged candidate sets ($|D|=50$).} $|\mathcal{C}|$ denotes the number of validated combinations present in the candidate set, AP corresponds the area under the precision-recall curve computed from the energy spectrum of the enlarged QUBOs, $P(k)$ is the precision at rank~$k$ from the exact energy ranking, $E(10)$ is the enrichment ratio at Rank $10$, and $f_{\text{match}}$ is the cumulative SQA frequency on validated combinations within the top~10 sampled configurations.}\label{tab:prediction_metrics}
\setlength{\tabcolsep}{3pt}
\renewcommand{\arraystretch}{1.4}
\begin{tabular}{l c c c c c c}
\toprule
Disease & $|\mathcal{C}|$ & AP & $P(5)$ & $P(10)$ & $E(10)$ & $f_{\text{match}}$ \\
\midrule
Diabetes Mellitus              & 17 & 0.456 & 0.80 & 0.60 & $734\times$ & 0.397 \\
\makecell[l]{Rheumatoid \\ Arthritis} & 11 & 0.399 & 0.80 & 0.40 & $757\times$ & 0.329 \\
Asthma                         & 11 & 0.408 & 0.60 & 0.40 & $757\times$ & 0.232 \\
Brain Neoplasms                & 12 & 0.376 & 0.80 & 0.40 & $694\times$ & 0.171 \\
\bottomrule
\end{tabular}
\end{table}

However, the disease-specific sampling distributions in Figure~\ref{fig:SQA} reveal notable structural differences.
Diabetes Mellitus and Rheumatoid Arthritis exhibit a somewhat dominant ground-state peak that concentrates over $20$\% of the total probability mass on a single validated combination ($f_1 = 0.261$ and $f_1 = 0.229$, respectively), with a sharp frequency drop to subsequent configurations. 
This behaviour is indicative of a well-separated spectral gap between the ground state and the first excited states (Figure~\ref{fig:spectrum}), which in turn reflects the large calibrated values of $\beta^*$ (13.39 and 14.47) that impose a strong energetic penalty per selected drug and thereby sharpen the energy landscape.

By contrast, Asthma displays a qualitatively different pattern. The second-ranked configuration is sampled more frequently than the ground state ($f_2 = 0.093$, $f_1 = 0.070$), and the top $10$ frequencies decay more gradually. 
This flatter distribution is consistent with the considerably smaller $\beta^* = 3.79$ and $\gamma^* = 1.46$, which produces a shallower energy landscape with tighter spectral gaps among the lowest-lying states.
In such a landscape, the SQA sampler distributes probability mass more uniformly across several near-degenerate configurations rather than concentrating on a single ground state pointing to a regime where the ability of the annealing process to harvest multiple near-optimal candidates becomes particularly advantageous.

Brain Neoplasms occupies an intermediate regime. While the ground-state peak is present ($f_1 = 0.081$), the overall top-$10$ frequency is lower ($\sum_{k=1}^{10} f_k = 0.214$), reflecting the broader energy scale induced by the largest $\beta^* = 34.38$ among the four diseases. 
The correspondingly wider spectral range means that, although the ground state is well separated in absolute energy, the relative probability assigned to it by the SQA sampler is diluted across a larger number of accessible low-energy states.

We present a brief literature-based evaluation of several top predicted combinations for each disease in Appendix~\ref{app:literature_validation}.

\section{Classical benchmarking and computational scaling}
\label{sec:benchmarking}

To assess the computational efficiency of the quantum annealing framework and contextualise it against classical optimisation, we benchmark the SQA sampler used throughout this work against a classical Simulated Annealing (SA) baseline. 
We compare both methods across multiple problem sizes and sweep budgets, using the validated drug combinations dataset as ground truth.

In order to establish a fair classical baseline, we employed the \texttt{SimulatedAnnealingSampler}~\cite{dwaveSimulatedAnnealingSampler} from the D-Wave Ocean SDK, which implements single-spin-flip Metropolis dynamics with a geometric inverse-temperature schedule interpolating between a specified $\beta_{\min}$ and $\beta_{\max}$. 

The SQA algorithm, which serves as the primary optimization tool throughout this study, was executed using the \texttt{PathIntegralAnnealingSampler}~\cite{dwavePathIntegralAnnealingSampler} from D-Wave's Ocean SDK.
This sampler implements SQA via path-integral Monte Carlo, where the transverse-field Ising model is mapped to a classical system with an additional imaginary-time dimension, so the system is represented by $P$ Trotter slices coupled along the Trotter direction, with the transverse field $\Gamma$ appearing as ferromagnetic coupling between adjacent slices. 
In this implementation, a sweep steps through the discretized-time schedule and updates spins according to a model-appropriate rule, so the work per sweep scales with the number of spin variables in the replicated system, typically $\mathcal{O}(nP)$, whereas SA performs one spin update per variable per temperature point ($\mathcal{O}(n)$) \cite{PhysRevB.78.134428}.
For the benchmarking comparison, both samplers use identical default geometric annealing schedules with the same inverse-temperature range $(\beta_{\min}, \beta_{\max}) = (0.1, 4.0)$ and the same number of Monte Carlo sweeps. 
 
We note that the discovery experiments presented in Section~\ref{sec:multiple_diseases} employ a separately tuned exponential annealing schedule for SQA, designed to promote exploration of near-degenerate low-energy states rather than to maximise the probability of returning the ground state. 
This schedule is not used in the benchmarking analysis, as it would artificially reduce the success probability for SQA and render the comparison unfair.

All experiments use $R = 1024$ independent reads per run averaged over multiple random subproblem draws to ensure statistical robustness.

To evaluate the computational efficiency of the solvers, we primarily rely on the time-to-solution (TTS) metric \cite{doi:10.1126/science.1252319}, which is derived from the empirical success probability of the sampling process.
The \emph{success probability} $p_s$ is the fraction of the $R$ independent reads that return any validated drug combination $c \in \mathcal{C}$:
\begin{equation}
    p_s = \frac{1}{R} \sum_{r=1}^{R} y_r,
\end{equation}
where $y_r$ is the indicator function defined in Eq.~\eqref{eq:indicator} for the combination returned by the $r$-th read.
Unlike a ground-state criterion, this definition is directly relevant to drug combination discovery since any validated combination constitutes a useful output, regardless of whether it corresponds to the global energy minimum.
The time-to-solution at confidence level $1 - \delta$ is a commonly employed metric in the benchmarking of QA experiments \cite{cao2016, boixo2014evidence}. It is defined as the total number of sweeps required to observe at least one validated combination with probability $\geq 1 - \delta$:
\begin{equation}
    \mathrm{TTS}_{1-\delta} = n_s \left\lceil
    \frac{\log \delta}{\log(1 - p_s)} \right\rceil,
    \label{eq:tts}
\end{equation}
where $n_s$ is the sweep budget, that is, the number of Monte Carlo sweeps performed per read. Throughout this work we use the standard choice of $\delta = 0.01$, yielding TTS$_{99}$ \cite{PhysRevE.108.065303, doi:10.1126/science.1252319}.
This metric is an accurate descriptor of the total algorithmic effort since a solver with higher $p_s$ at the same sweep budget will have lower TTS as fewer independent reads are needed to guarantee success with high confidence. 
When $p_s = 0$, TTS is undefined.

The uncertainty in TTS is propagated from the measured variance in $p_s$ via
\begin{equation}
    \sigma_{\mathrm{TTS}} \approx \left|
    n_s \frac{\log \delta}{(1 - p_s)\,[\log(1 - p_s)]^2}
    \right| \sigma_{p_s},
\end{equation}
obtained by first-order error propagation of Eq.~\eqref{eq:tts}.

\begin{figure*}[htbp]
  \centering

  \begin{subfigure}[b]{0.48\textwidth}
    \refstepcounter{subfigure}\label{fig:subttsa}%
    \begin{tikzpicture}
      \node[inner sep=0pt] (img) {\includegraphics[width=\linewidth]{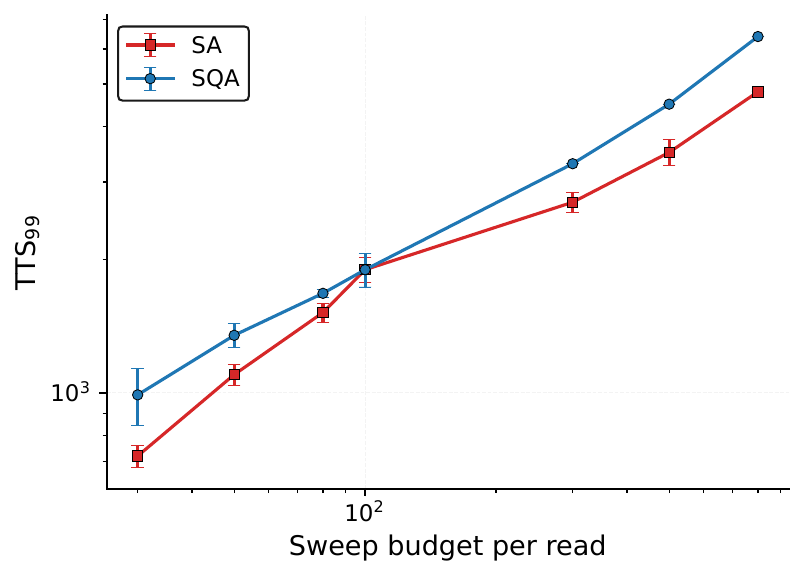}};
      \node[anchor=north west, yshift=10pt,fill=white,inner sep=2pt,font=\bfseries] at (img.north west) {(\thesubfigure)};
    \end{tikzpicture}
  \end{subfigure}\hfill
  \begin{subfigure}[b]{0.48\textwidth}
    \refstepcounter{subfigure}\label{fig:subttsb}%
    \begin{tikzpicture}
      \node[inner sep=0pt] (img) {\includegraphics[width=\linewidth]{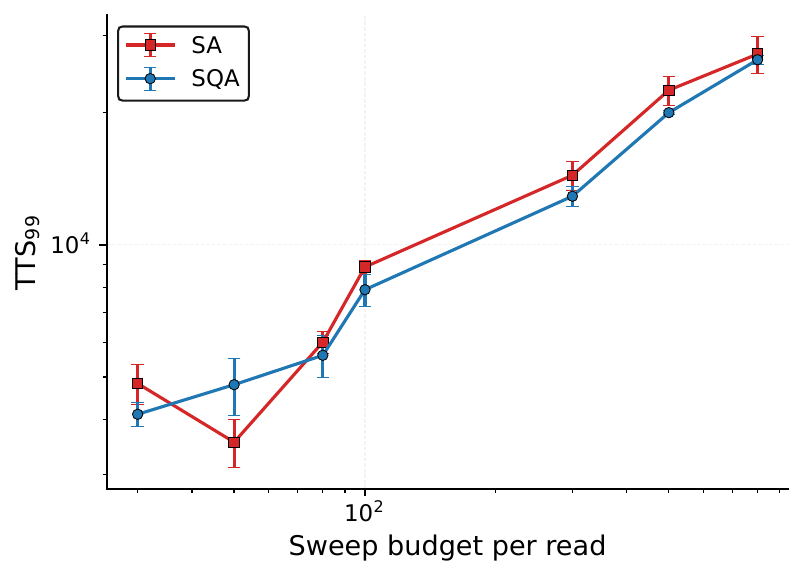}};
      \node[anchor=north west, yshift=10pt,fill=white,inner sep=2pt,font=\bfseries] at (img.north west) {(\thesubfigure)};
    \end{tikzpicture}
  \end{subfigure}

  \vspace{2mm}

  \begin{subfigure}[b]{0.48\textwidth}
    \refstepcounter{subfigure}\label{fig:subttsc}%
    \begin{tikzpicture}
      \node[inner sep=0pt] (img) {\includegraphics[width=\linewidth]{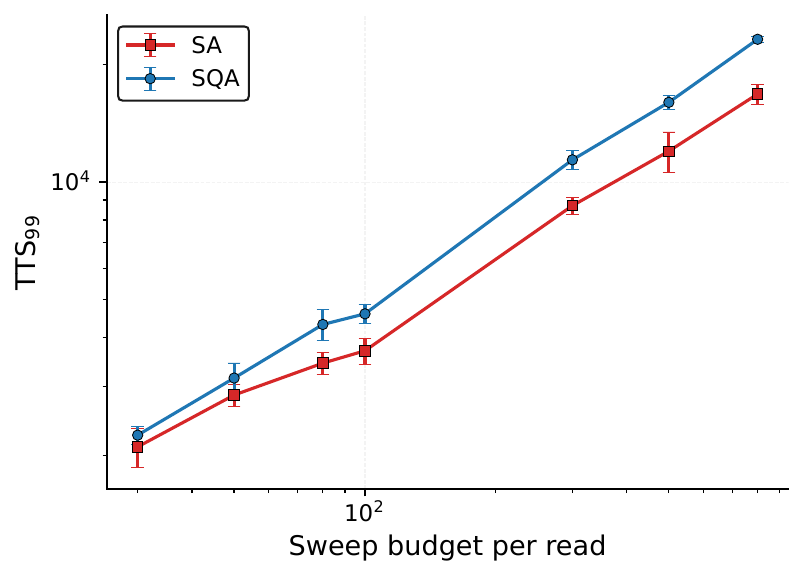}};
      \node[anchor=north west, yshift=10pt,fill=white,inner sep=2pt,font=\bfseries] at (img.north west) {(\thesubfigure)};
    \end{tikzpicture}
  \end{subfigure}\hfill
  \begin{subfigure}[b]{0.48\textwidth}
    \refstepcounter{subfigure}\label{fig:subttsd}%
    \begin{tikzpicture}
      \node[inner sep=0pt] (img) {\includegraphics[width=\linewidth]{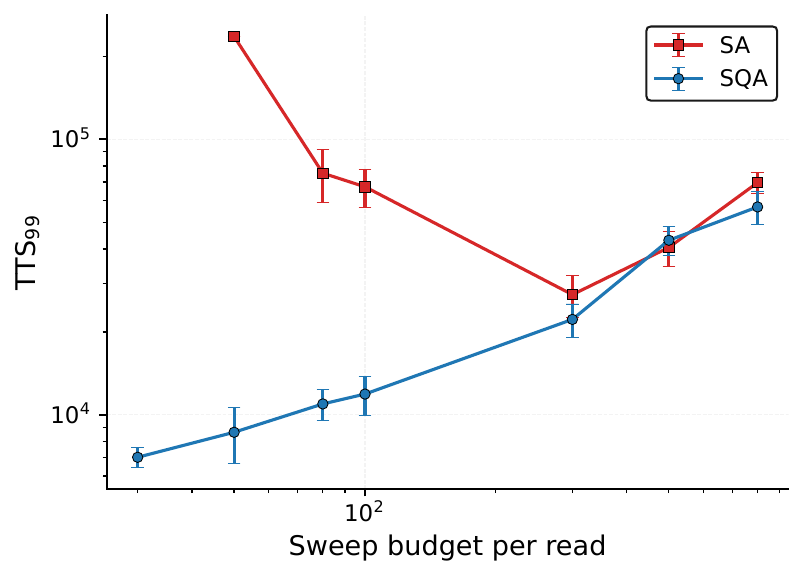}};
      \node[anchor=north west, yshift=10pt,fill=white,inner sep=2pt,font=\bfseries] at (img.north west) {(\thesubfigure)};
    \end{tikzpicture}
  \end{subfigure}

  \caption{\justifying \textbf{TTS$_{99}$ as a function of sweep budget} at fixed problem size $n = 50$ for SA (red) and SQA (blue). 
  Both solvers use identical default geometric annealing schedules with 1024 reads. 
  The non-monotonic behaviour reflects the trade-off between increasing $p_s$ (which lowers the number of required reads) and increasing per-read cost $n_s$. 
  Error bars denote the propagated uncertainty from $\sigma_{p_s}$. The diseases represented are: \textbf{(a)} Diabetes Mellitus, \textbf{(b)} Rheumatoid Arthritis, \textbf{(c)} Asthma, and \textbf{(d)} Brain Neoplasms.}
\label{fig:tts_sweeps}
\end{figure*}

Before evaluating how the problem scales with the number of candidate drugs, we first examine how TTS depends on the sweep budget $n_s$ at fixed problem size ($n = 50$) for each disease. The results are shown in Figure~\ref{fig:tts_sweeps}.

The TTS curves sometimes exhibit non-monotonic behaviour. 
At low sweep budgets, $p_s$ is small and TTS is dominated by the $\lceil \log\delta / \log(1-p_s) \rceil$ factor, which diverges as $p_s \to 0$.
As $n_s$ increases, $p_s$ grows and TTS initially decreases. 
However, beyond a certain point, further increases in $n_s$ yield diminishing returns in $p_s$ while the per-read cost $n_s$ continues to grow linearly, causing TTS to increase again. 
This is particularly clear for Brain Neoplasms whose optimal sweep budget (the value of $n_s$ minimising TTS) falls in the range $n_s \approx 300$--$500$.
This analysis motivates the choice of a fixed sweep budget of $n_s =500$ for the scaling experiments. At this value, both solvers operate well above their optimal sweep budget, in a regime where $p_s$ has largely saturated. 
This choice is conservative and, if anything, slightly favours SA.

To study how computational cost grows with problem size, we construct subproblems of sizes $n$ ranging from $10$ to $50$ by drawing random subsets from the $50$-drug enlarged candidate pools. 
For each size, the core drugs from validated combinations are always included, and the remainder are randomly sampled from the padding pool with results averaged over multiple random draws.

\begin{figure*}[htbp]
  \centering

  \begin{subfigure}[b]{0.48\textwidth}
    \refstepcounter{subfigure}\label{fig:subttsfita}%
    \begin{tikzpicture}
      \node[inner sep=0pt] (img) {\includegraphics[width=\linewidth]{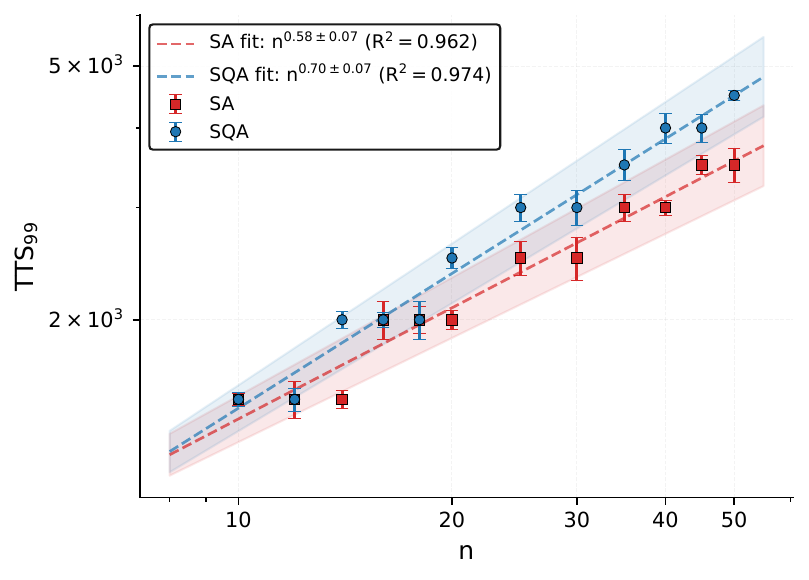}};
      \node[anchor=north west, yshift=10pt,fill=white,inner sep=2pt,font=\bfseries] at (img.north west) {(\thesubfigure)};
    \end{tikzpicture}
  \end{subfigure}\hfill
  \begin{subfigure}[b]{0.48\textwidth}
    \refstepcounter{subfigure}\label{fig:subttsfitb}%
    \begin{tikzpicture}
      \node[inner sep=0pt] (img) {\includegraphics[width=\linewidth]{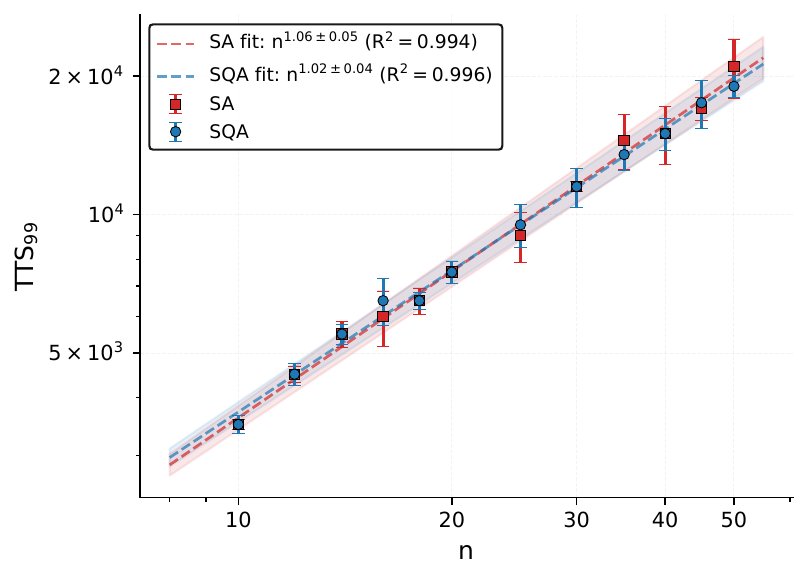}};
      \node[anchor=north west, yshift=10pt,fill=white,inner sep=2pt,font=\bfseries] at (img.north west) {(\thesubfigure)};
    \end{tikzpicture}
  \end{subfigure}

  \vspace{2mm}

  \begin{subfigure}[b]{0.48\textwidth}
    \refstepcounter{subfigure}\label{fig:subttsfitc}%
    \begin{tikzpicture}
      \node[inner sep=0pt] (img) {\includegraphics[width=\linewidth]{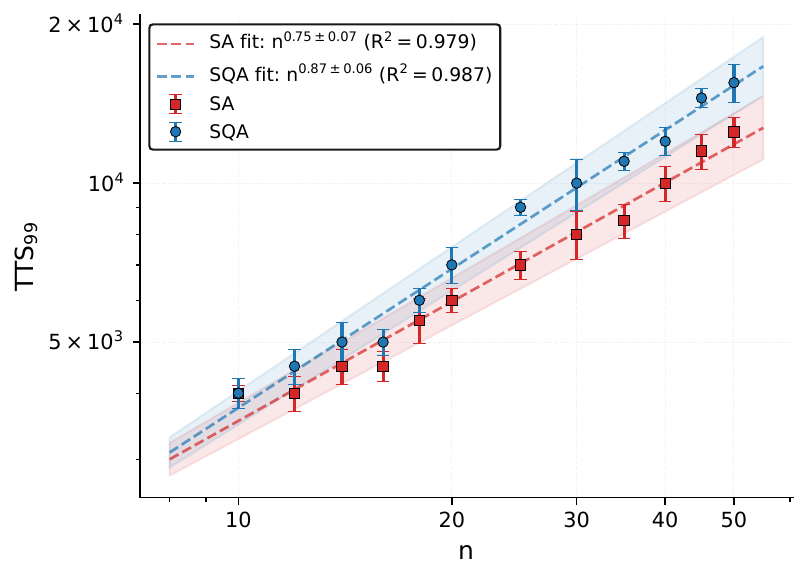}};
      \node[anchor=north west, yshift=10pt,fill=white,inner sep=2pt,font=\bfseries] at (img.north west) {(\thesubfigure)};
    \end{tikzpicture}
  \end{subfigure}\hfill
  \begin{subfigure}[b]{0.48\textwidth}
    \refstepcounter{subfigure}\label{fig:subttsfitd}%
    \begin{tikzpicture}
      \node[inner sep=0pt] (img) {\includegraphics[width=\linewidth]{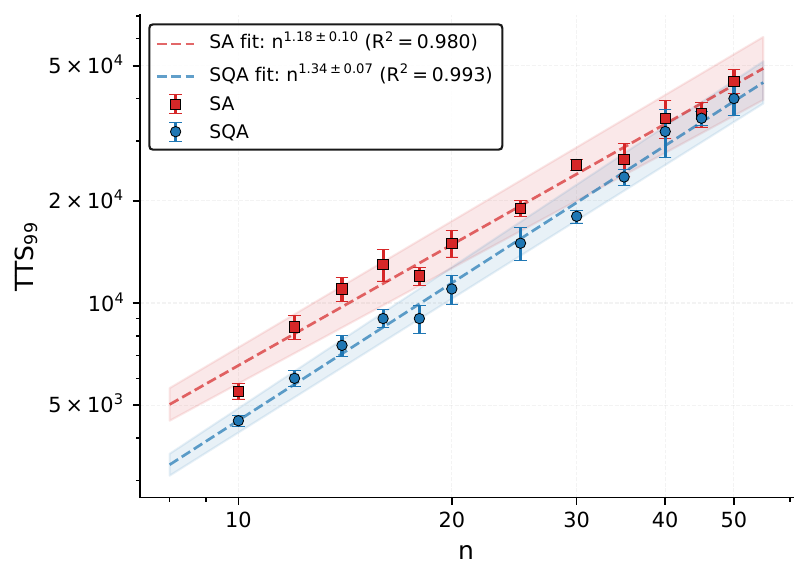}};
      \node[anchor=north west, yshift=10pt,fill=white,inner sep=2pt,font=\bfseries] at (img.north west) {(\thesubfigure)};
    \end{tikzpicture}
  \end{subfigure}

\caption{\justifying \textbf{TTS$_{99}$ scaling with problem size.} Each panel shows TTS$_{99}$ as a function of the number of candidate drugs $n$ for SA (red) and SQA (blue) on a log-log scale, at a fixed sweep budget of $n_s = 500$. 
  Dashed lines show polynomial fits $\mathrm{TTS} \sim n^{\alpha}$, with the fitted exponents, $95$\% confidence intervals, and $R^2$ values indicated in the legends.
  Shaded regions denote the uncertainty band on $\alpha$. Subproblems are constructed by drawing random subsets from the $50$-drug enlarged candidate pool, always including the core drugs from validated combinations. 
  Error bars denote propagated uncertainty from $\sigma_{p_s}$. The diseases represented are: \textbf{(a)} Diabetes Mellitus, \textbf{(b)} Rheumatoid Arthritis, \textbf{(c)} Asthma, and \textbf{(d)} Brain Neoplasms.}
  \label{fig:tts_scaling}
\end{figure*}

The TTS scaling results for all four diseases are summarised in Figure~\ref{fig:tts_scaling}, where the per-disease curves and their best fits are displayed.
We fit both a polynomial model $\mathrm{TTS} \sim n^{\alpha}$ and an exponential model $\mathrm{TTS} \sim e^{cn}$ to the data and report both fits with $95$\% confidence intervals in Table~\ref{tab:scaling}. 
Comparing the results, we find a scaling empirically consistent with polynomial growth in the range of problem sizes studied. 
We note, however, that the accessible problem sizes do not allow us to conclusively rule out a slowly growing exponential and a definite comparative assessment of the performance between the two models would require data at substantially larger~$n$. 

\begin{figure*}[htbp]
  \centering

  \begin{subfigure}[b]{0.48\textwidth}
    \refstepcounter{subfigure}\label{fig:subttsscaling_sa}%
    \begin{tikzpicture}
      \node[inner sep=0pt] (img) {\includegraphics[width=\linewidth]{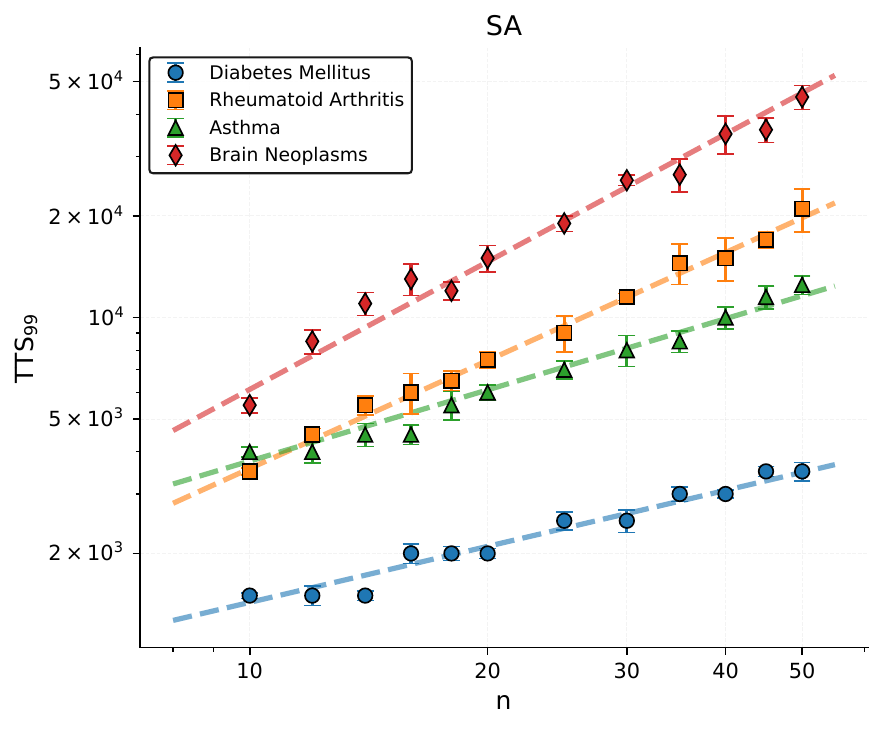}};
      \node[anchor=north west, yshift=10pt,fill=white,inner sep=2pt,font=\bfseries] at (img.north west) {(\thesubfigure)};
    \end{tikzpicture}
  \end{subfigure}\hfill
  \begin{subfigure}[b]{0.48\textwidth}
    \refstepcounter{subfigure}\label{fig:subttsscaling_sqa}%
    \begin{tikzpicture}
      \node[inner sep=0pt] (img) {\includegraphics[width=\linewidth]{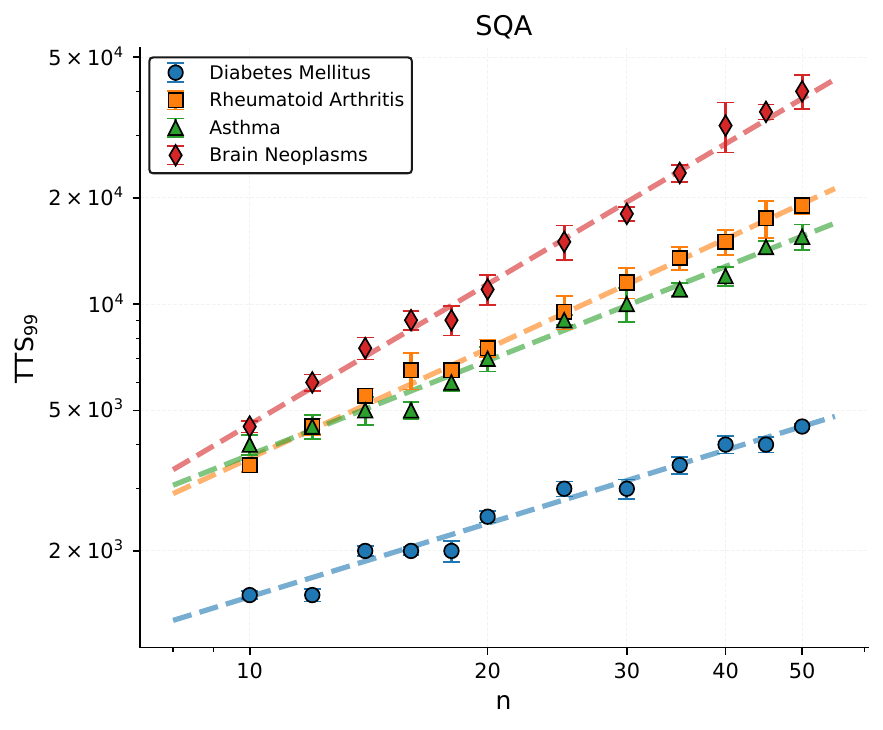}};
      \node[anchor=north west, yshift=10pt,fill=white,inner sep=2pt,font=\bfseries] at (img.north west) {(\thesubfigure)};
    \end{tikzpicture}
  \end{subfigure}

  \caption{\justifying \textbf{Scaling comparison.}
  Each panel displays the polynomial scaling of the Time-to-Solution ($\mathrm{TTS}_{99}$) with the number of candidate drugs $n$ for \textbf{(a)} Simulated Annealing (SA) and \textbf{(b)} Simulated Quantum Annealing (SQA) on a log-log scale across all four diseases. The steeper scaling observed for Brain Neoplasms and Rheumatoid Arthritis ($\alpha \gtrsim 1$) in both solvers, compared to Diabetes Mellitus and Asthma ($\alpha < 0.9$), reflects the greater computational difficulty of their respective QUBO landscapes. This increased difficulty correlates with the lower Average Precision values reported in Table~\ref{tab:hyperparams}.}
\label{fig:tts_all}
\end{figure*}

This varying relative performance of SA and SQA across diseases, illustrated more directly in the cross-disease comparison of Figure~\ref{fig:tts_all}, is reflective of the differences in the structure of the underlying QUBO energy landscapes. 
To assess whether the observed differences are meaningful, we compare the fitted polynomial exponents and their $95$\% confidence intervals.

For Diabetes Mellitus, SA achieves higher $p_s$ than SQA at all problem sizes (e.g.\ $p_s = 0.494 \pm 0.022$ vs.\ $0.422 \pm 0.006$ at $n = 50$) and correspondingly lower TTS. 
The fitted scaling exponents are $\alpha_{\mathrm{SA}} = 0.58 \pm 0.07$ and $\alpha_{\mathrm{SQA}} = 0.70 \pm 0.07$, whose confidence intervals show only marginal overlap. 
This would indicate that it is likely for a slight advantage in favor of SA for this disease, which can be understood from the well-separated energy spectrum shown in Figure~\ref{fig:spectrum}a. 
The large calibrated $\beta^* = 13.39$ produces a pronounced spectral gap, creating a smooth energy landscape that classical thermal fluctuations already navigate efficiently. 
Therefore, in this regime, the overhead of simulating Trotter replicas in SQA likely does not confer a compensating advantage.

Regarding Rheumatoid Arthritis, SA and SQA perform nearly identically at all problem sizes, with $p_s$ values differing by less than $0.01$ in most cases and TTS curves that are statistically indistinguishable.
The scaling exponents are essentially equal ($\alpha_{\mathrm{SA}} = 1.06 \pm 0.05$ and $\alpha_{\mathrm{SQA}} = 1.02 \pm 0.04$), which suggests that, for this disease, the energy landscape presents comparable difficulty to both thermal and tunnelling-based exploration.

For Asthma, SA maintains moderately higher $p_s$ than SQA across most problem sizes, with corresponding scaling exponents $\alpha_{\mathrm{SA}} = 0.75 \pm 0.07$ and $\alpha_{\mathrm{SQA}} = 0.87 \pm 0.06$. 
The confidence intervals again show only marginal overlap, which while consistent with a mild SA advantage, should be interpreted with caution given the limited range of problem sizes. 
Interestingly, the relatively small $\beta^* = 3.79$ calibrated for Asthma is expected to produce a shallower energy landscape with tighter spectral gaps among low-lying states. 
While this structure is easily navigated by quantum tunnelling, SQA does not seem to yield a competitive advantage, likely because the thermal hopping of SA is already sufficient to overcome the barriers of the flatter energy landscape.

\begin{table*}[htbp]
\centering
\caption{\justifying \textbf{Scaling fits.} Best-fit exponents for polynomial ($\mathrm{TTS} \sim n^{\alpha}$) and exponential ($\mathrm{TTS} \sim e^{cn}$) models for SA and SQA across all four diseases. Uncertainties denote $95$\% confidence intervals from the least-squares fit.}
\label{tab:scaling}
\setlength{\tabcolsep}{8pt}
\renewcommand{\arraystretch}{1.25}
\begin{tabular}{@{} l cccc cccc @{}}
\toprule
& \multicolumn{4}{c}{\textbf{SA}} & \multicolumn{4}{c}{\textbf{SQA}} \\
\cmidrule(lr){2-5} \cmidrule(l){6-9}
Disease 
& \multicolumn{2}{c}{Polynomial} & \multicolumn{2}{c}{Exponential} 
& \multicolumn{2}{c}{Polynomial} & \multicolumn{2}{c}{Exponential} \\
\cmidrule(lr){2-3} \cmidrule(lr){4-5} \cmidrule(lr){6-7} \cmidrule(l){8-9}
& $\alpha$ & $R^2$ & $c$ & $R^2$ 
& $\alpha$ & $R^2$ & $c$ & $R^2$ \\
\midrule
Diabetes Mellitus    & $0.58 \pm 0.07$ & $0.962$ & $0.023 \pm 0.004$ & $0.929$ & $0.70 \pm 0.07$ & $0.974$ & $0.027 \pm 0.005$ & $0.927$ \\
Rheumatoid Arthritis & $1.06 \pm 0.05$ & $0.994$ & $0.041 \pm 0.005$ & $0.959$ & $1.02 \pm 0.04$ & $0.996$ & $0.040 \pm 0.006$ & $0.947$ \\
Asthma               & $0.75 \pm 0.07$ & $0.979$ & $0.030 \pm 0.003$ & $0.975$ & $0.87 \pm 0.06$ & $0.987$ & $0.034 \pm 0.005$ & $0.957$ \\
Brain Neoplasms      & $1.18 \pm 0.10$ & $0.980$ & $0.046 \pm 0.008$ & $0.926$ & $1.34 \pm 0.07$ & $0.993$ & $0.053 \pm 0.006$ & $0.968$ \\
\bottomrule
\end{tabular}
\end{table*}

Brain Neoplasms is the only disease for which SQA consistently achieves higher $p_s$ than SA. 
For instance, at $n = 14$, SQA achieves $p_s = 0.280 \pm 0.018$ compared to $p_s = 0.193 \pm 0.014$ for SA, representing a 45\% relative improvement.
However, the scaling exponents $\alpha_{\mathrm{SA}} = 1.18 \pm 0.10$ and $\alpha_{\mathrm{SQA}} = 1.34 \pm 0.07$ show yet again marginally overlapping confidence intervals. 
Brain Neoplasms has the largest calibrated $\beta^* = 34.38$ of all studied diseases, which produces the widest absolute energy scale and, consequently, the most rugged landscape with deep local minima. 
This could be indicative of a regime where quantum tunnelling provides an advantage over thermal hopping, as the system can tunnel through energy barriers that SA must climb over~\cite{das2008}. 
The convergence of TTS values at large $n$ ($n \geq 45$) could also reflect the limitations of simulated tunnelling where as the problem grows, the Trotter decomposition becomes an increasingly coarse approximation of true quantum dynamics, diluting the tunnelling advantage \cite{PhysRevLett.117.180402}.

\section{Discussion}

We have introduced a network-medicine informed quantum annealing pipeline that encodes the Complementary Exposure principle as a QUBO and uses low-energy configurations to prioritise candidate drug combinations. 
The results, obtained for the four exemplar diseases of Diabetes Mellitus, Rheumatoid Arthritis, Asthma, and Brain Neoplasms, demonstrate that low-energy states of the constructed Hamiltonian are considerably populated with validated combinations, indicating that topological proxies such as network proximity and separation carry useful information for combination discovery.

Because our algorithm is driven exclusively by graph descriptors, its outputs should be read as ranked hypotheses rather than mechanistic or clinical proof. 
A low QUBO energy mainly identifies candidate combinations with high network proximity to the disease module while maintaining minimal target overlap, a correlative signal, not direct evidence of effectiveness or safety.
Additionally, the hyperparameters that balance the cardinality of the combinations ($\beta$) and pairwise interactions ($\gamma$) were calibrated against a constructed benchmark of validated combinations and found to vary reasonably across the four diseases analysed. 
This dependence highlights that the fitted values are contingent on the composition and representativity of the benchmarking set. Thus, with small or skewed datasets, the calibrated hyperparameters are more likely to reflect dataset-specific properties than in transferable, disease-agnostic settings.

Relatedly, the Complementary Exposure principle itself was originally formulated and validated in restricted biological contexts (notably hypertension and cancer) and for drug pairs only. There is no a priori guarantee that the same topological signature of complementarity will hold universally across the entire disease spectrum. Accordingly, extending and applying the principle to new disease areas should be performed with per-disease validation.

The results across the four diseases studied offer a nuanced view of how broadly the Complementary Exposure principle transfers. 
For Diabetes Mellitus and Rheumatoid Arthritis, the principle appears to hold robustly. 
The calibrated Hamiltonians achieve the highest AP values ($0.670$ and $0.594$, respectively) and the energy spectra show tight clustering of validated combinations at low energies (Figure~\ref{fig:spectrum}). 
These are diseases with large, well-characterised disease modules and relatively rich combination datasets, providing favorable conditions for the topological signals encoded in the $z$-score and separation measure to be informative.
The brief literature search conducted in Appendix~\ref{app:literature_validation} further supports this. For Rheumatoid Arthritis, the algorithm independently identifies the adalimumab--methotrexate combination, which a large randomised clinical trial found to achieve approximately double the remission rate of either drug alone~\cite{breedveld2006premier}. Additionally, for Diabetes Mellitus, the top-ranked prediction (metformin--irbesartan) corresponds to a pairing whose additive protective effect has been demonstrated in a laboratory setting~\cite{ishibashi2012beneficial}.

The picture is more complex for Asthma and Brain Neoplasms. 
Asthma presents the unusual case where all candidate drugs, including those from validated combinations, exhibit $z > 0$ (Figure~\ref{fig:zs_distributions}), indicating that the strict $z < 0$ criterion of the original Complementary Exposure formulation is not met. 
Yet the framework still achieves AP~$= 0.600$ on the calibration set. 
This suggests that the relative ordering of $z$-scores, rather than their absolute sign, carries useful information, and that the hyperparameter $\beta$ can compensate for uniformly positive $z$ values by adjusting the selection threshold. 
However, it also raises the question of whether, for certain diseases, the interactome-based proximity measure may be systematically biased. 
For instance, if the disease module is located in a region of the interactome that is particularly incomplete or where the degree distribution departs from the assumptions of the null model. As noted in Appendix~\ref{app:literature_validation}, several of the Asthma predictions involve arformoterol, a drug approved for COPD rather than asthma, which may point to network proximity between the two disease modules rather than asthma-specific therapeutic relevance.

Brain Neoplasms reveals a different limitation.
Although the calibration AP ($0.528$) is acceptable and the SQA sampler recovers validated combinations at enrichment ratios exceeding $694\times$ (Table~\ref{tab:prediction_metrics}), a brief literature search (Appendix~\ref{app:literature_validation}) reveals that several of the recurring predicted drugs (vinorelbine, epirubicin, tamoxifen, and capecitabine) appear predominantly in the context of breast cancer treatment rather than brain tumours.
This pattern could indicate that the algorithm is capturing shared oncogenic network signatures rather than brain-specific therapeutic relevance, though a more careful assessment would be required by experts. 
This observation, together with the fact that Brain Neoplasms yields the lowest AP among the four diseases, suggest that the quality and disease-specificity of the benchmarking data plays a decisive role in determining whether the framework's outputs are clinically meaningful. 
Furthermore, it raises the possibility that the complementary exposure principle itself may be less fundamentally applicable to the network topology of brain neoplasms than to the other conditions studied.

These remarks point to a broader concern about overfitting to the benchmark dataset.
The hyperparameters $(\gamma^*, \beta^*)$ are calibrated by maximising AP on the validated combinations $\mathcal{C}$, which for some diseases comprises as few as 11 ground-truth positives drawn from a candidate pool of $165$ configurations.
In this low-data regime, the calibrated parameters risk capturing statistical artefacts of the specific drug pool rather than transferable properties of the disease biology. 
The variation in $\beta^*$ across diseases, ranging from $3.79$ (Asthma) to $34.38$ (Brain Neoplasms), may partly reflect such overfitting, as these values are sensitive to the composition of $\mathcal{C}$ and the particular drugs it contains. 
The partial mitigation observed in our experiments, where calibrated parameters generalise to $50$-drug enlarged sets with significant enrichment ratios, is encouraging but does not rule out overfitting to structural features shared between the calibration and enlarged pools, since the core drugs from $\mathcal{C}$ are always present in both.

When curated benchmark datasets are unavailable or sparse, as is the case for rarer diseases, newly characterised conditions, or higher-order combinations, the calibration procedure described in Section~\ref{sec:calibration} may also not be directly applicable. 
In such scenarios, one possible strategy is to transfer hyperparameters from a related disease with a richer combination dataset, leveraging the observation that diseases with similar interactome topology tend to share pharmacological characteristics, however this generalisation is yet to be tested. 
Alternatively, the hyperparameters could be set to disease-agnostic defaults derived from cross-disease averages, accepting a reduction in AP in exchange for broader applicability. 
A more principled approach would be to develop regularisation or cross-validation strategies that explicitly penalise overfitting to small calibration sets, for instance by holding out a fraction of the validated combinations during the grid search and evaluating generalisation on the held-out set. We opted not to do this, due to the limited availability of large-scale validation datasets. 
Exploring these directions is essential for extending the framework beyond the studied diseases considered here.

A related limitation concerns the quality and coverage of the underlying protein-protein interactome.
The network used in this work, while one of the most comprehensive experimentally grounded map available, captures an estimated $25$--$35$\% of all true physical interactions~\cite{Stumpf2008,Rolland2014} and is subject to investigative biases that favour well-studied proteins and pathways~\cite{Rolland2014}. 
These issues are not unique to our framework but affect all network medicine approaches that rely on interactome data. 
Importantly, Menche et al.~\cite{Menche2015} conducted explicit robustness tests by repeating their disease module analysis using only unbiased high-throughput yeast-two-hybrid data, discarding all literature-curated interactions, and found that the key network measure such as module diameters ($\langle d_{AA} \rangle$), inter-module distances ($\langle d_{AB} \rangle$), and separation values $s(A,B)$ remain highly significantly correlated between the full and unbiased interactomes.
An analogous test replacing curated disease-gene lists with unbiased genome-wide association data yielded the same conclusion.
Furthermore, the consistency of the separation measure across interactome versions of substantially different size, from the 141,296-edge network of Menche et al.\ to the 243,603-edge network of Cheng et al.~\cite{Cheng2019} used here, provides indirect evidence that the network proximity signals exploited by our Hamiltonian are robust to incremental improvements in interactome coverage. 
Nevertheless, a systematic sensitivity analysis using perturbed or alternative network constructions would further strengthen confidence in the generalisability of the approach and constitutes a natural direction for future work.

The benchmarking analysis presented in Section~\ref{sec:benchmarking} provides a conservative assessment of the computational landscape. 
Across all four diseases, the observed TTS scaling is empirically consistent with polynomial growth in the range of problem sizes studied, though the accessible sizes ($n \leq 50$) do not permit conclusive rejection of a slowly growing exponential. 
Additionally, the classical SA baseline appears to match or outperform SQA on three of the four diseases when both are given identical schedules and sweep budgets. 
The exception would be Brain Neoplasms, which is characterized by the highest calibrated $\beta^*$ and, consequently, the widest energy scale and most rugged energy landscape. In this case, SQA achieves up to $45$\% higher success probability at intermediate problem sizes, an observation consistent with the theoretical expectation that quantum tunnelling provides the greatest benefit in landscapes with tall but thin energy barriers~\cite{das2008,AnnealingReview,PhysRevX.6.031015}.
This advantage diminishes at larger sizes, likely due to the Trotter decomposition becoming an increasingly coarse approximation of true quantum dynamics. 

These results should be interpreted with care as SQA is a classical simulation of quantum annealing, not a substitute for it. 
On physical quantum annealing hardware, tunnelling occurs through genuine quantum coherence rather than through classical Monte Carlo moves across Trotter slices, and the per-anneal time is determined by the physical annealing schedule (typically microseconds) rather than by the number of algorithmic sweeps~\cite{AnnealingReview,10.14778/2947618.2947621}. 
Whether the apparent landscape-dependent advantage observed for Brain Neoplasms translates into a genuine quantum speedup on hardware remains an open experimental question.

The scaling consistent with polynomial growth observed here is, however, a necessary condition for quantum annealers to offer practical utility on these instances~\cite{knysh2016zero}.

It is also worth noting that while strict benchmarking metrics like TTS are necessary to assess algorithmic tractability, the ultimate utility of this framework also depends on solution diversity.
A sampler's ability to broadly explore the low-energy landscape and return a diverse set of novel combinations is arguably as important as its ability to quickly locate a single validated ground state.

Furthermore, while our construction can be restricted to pairwise combinations, the Hamiltonian formulation naturally extends to candidate sets of arbitrary size. 
In practice, however, the results presented in this work are overwhelmingly dominated by pairwise predictions.
For example, all top-five predictions across the four diseases correspond to drug pairs rather than triplets, despite the constraint (Eq.~\ref{eq:dbl_constraint}) admitting both $k=2$ and $k=3$. 
This reflects both the composition of the benchmarking dataset $\mathcal{C}$, in which 234 of 287 validated combinations are pairwise, and the correspondingly large calibrated $\beta^*$ values that impose a steep energetic cost per additional selected drug. 
Rigorously evaluating higher-order predictions would require validated combination datasets of three or more drugs that are currently scarce (only 40 triplets and 13 combinations of four or more drugs appear in $\mathcal{C}$ across all 35 diseases) and whose network-proximity statistics are less well characterised in the literature, as the Complementary Exposure principle was originally formulated and validated for drug pairs only~\cite{Cheng2019}.
Extending both the theoretical framework and the empirical validation to higher-order combinations remains an important open direction.

Likewise, the interpretation of the quadratic coupling parameter $\gamma$ is flexible. While positive $\gamma$ emphasises complementary and non-overlapping targeting, scenarios where $\gamma$ takes negative values could be explored to prioritise antagonistic interactions by consequence of overlapping exposure.
In addition, the annealing paradigm naturally returns a distribution over many low-energy configurations rather than a single optimum, which is a strength in discovery settings, allowing harvesting of multiple near-optimal combination hypotheses from an astronomically large space.

Taken together, the quantum-annealing formulation we propose does not replace biological experimentation, rather, it narrows the combinatorial search space to a manageable, ranked list of novel hypotheses that can be prioritised for experimental testing.
When used in this way, the framework can speed up the early stages of combination discovery.

If embedded  within a rigorous experimental validation loop and complemented by pharmacological modelling, this approach could contribute to accelerating the identification of biologically plausible drug combinations and help focus resources on the most promising hypotheses.

\subsection*{Acknowledgements}
The authors would like to thank Lorenzo Buffoni for useful discussions on quantum annealing.

\subsection*{Funding}
This work was supported by FCT - Fundação para a Ciência e Tecnologia trough project UIDB/50008/2020, with the DOI identifier 10.54499/UIDB/50008/2020, trough project QuNetMed 2022.05558.PTDC, with the DOI identifier 10.54499/2022.05558.PTDC.

\subsection*{Author Contributions}
B. C. conceived the original idea of using quantum annealing for drug combination discovery. 
D. R. and D.M. developed the quantum annealing framework.
D. R. prepared the datasets and performed the numerical simulations.
All authors contributed to the writing and revision of the manuscript.

\subsection*{Data and Code Availability}
All code required to reproduce the analyses, figures and experiments in this work is publicly available at \url{https://github.com/dmrapk/Drug-Combinations-using-Quantum-Annealers}.
The validated benchmark disease-drug-combination dataset compiled for this study as well as an extended list of the predicted combinations is included in that repository. All other datasets used in the study (interactome, drug-targets and disease-gene associations) are publicly available and their original sources are listed and cited in Appendix \ref{sec:datasets}.

\appendix

\section{Datasets}
\label{sec:datasets}
\subsection{Human Protein-Protein Interactome}

The underlying human protein-protein interactome is modelled as an undirected, unweighted graph $\mathcal{G} = (V, E)$, where each node $v \in V$ represents a protein and each edge $(u,v) \in E$ denotes a documented physical interaction. We utilize the high-quality interactome compiled by Cheng et al.~\cite{Cheng2019} (Supplementary Data~1). The original authors constructed this network by aggregating data from fifteen distinct biomedical databases, removing duplicated pairs, and standardizing protein-coding genes to official gene symbols and Entrez IDs via the NCBI database~\cite{NCBI2016} and GeneCards (\url{http://www.genecards.org/}). 

To guarantee biological relevance and minimize false positives, the Cheng \textit{et al.} dataset restricts its edges strictly to interactions supported by direct experimental evidence. Specifically, an edge $(u,v)$ is included in their network only if it is corroborated by at least one of five lines of evidence: (i) binary interactions from high-throughput yeast-two-hybrid (Y2H) screens~\cite{Rolland2014,Rual2005} and an unpublished CCSB dataset (\url{https://ccsb.dana-farber.org/interactome-data.html}); (ii) kinase-substrate interactions derived from phosphoproteomics; (iii) literature-curated interactions identified via affinity purification-mass spectrometry (AP-MS), Y2H, or low-throughput methods; (iv) binary interactions resolved from 3D protein structures; and (v) directed physical interactions within established signaling cascades. Crucially, any computationally inferred association were excluded leaving only experimentally validated correspondences. The reliability of this interactome is derived from the strict quality-control protocols of its primary data sources. For instance, the binary physical interactions from high-throughput yeast-two-hybrid (Y2H) screens were subjected to retesting protocols that maintained a verified false-positive rate of approximately 5\%~\cite{Rolland2014}. Similarly, interactions identified via affinity purification-mass spectrometry (AP-MS) were typically filtered using rigorous statistical scoring (such as CompPASS~\cite{Sowa2009} or SAINT~\cite{Choi2011}) to ensure a False Discovery Rate (FDR) of $\leq 0.05$. 

The resulting provided graph comprises $|V| = 16{,}677$ nodes and $|E| = 243{,}603$ edges. Despite this, current interactome maps remain highly incomplete. Based on theoretical and empirical projections that place the true size of the human interactome between 650,000 and 1,000,000 physical interactions~\cite{Stumpf2008,Rolland2014}, this network is estimated to capture only roughly 25--35\% of all true pairwise interactions in the human cell. 
This inherent incompleteness has considerable consequences for the detectability of disease modules.
Consequently, this missing data triggers a percolation-like fragmentation~\cite{Menche2015}, causing disease modules that are continuous in the cell to shatter into isolated subgraphs within the mapped interactome. Menche \textit{et al.}~\cite{Menche2015} demonstrated that below a critical module size, this fragmentation makes it impossible to reliably distinguish a disease's localized network neighborhood from random background noise. Consequently, we restrict our analysis to diseases with a substantial number of mapped genes, ensuring their network neighborhoods remain coherent enough to yield reliable shortest-path measurements.

\subsection{Drug-target interactions}

Drug-target interactions were taken from the dataset compiled by Cheng, Kov\'acs \& Barab\'asi~\cite{Cheng2019} (Supplementary Data~2), which aggregates high-quality binding affinity profiles from six data sources: DrugBank~\cite{Law2014}, the Therapeutic Target Database (TTD)~\cite{Zhu2012}, PharmGKB~\cite{Hernandez2008}, ChEMBL~\cite{Gaulton2012}, BindingDB~\cite{Liu2007}, and the IUPHAR/BPS Guide to PHARMACOLOGY~\cite{Pawson2014}. A drug-target pair was retained only if it satisfied four criteria: a reported binding affinity, demonstrated by inhibition constant ($K_i$), dissociation constant ($K_d$), median effective concentration (EC$_{50}$), or median inhibitory concentration (IC$_{50}$) of $\leq 10\;\mu$M; a unique UniProt accession number; a ``reviewed'' status in the UniProt database~\cite{UniProt2019}; and \textit{Homo sapiens} origin. After applying these filters, the dataset comprises 15,051 drug-target interactions connecting 4,428 drugs to 2,256 unique human proteins, of which 1,978 drugs have at least two experimentally validated targets and with an average number of targets per drug of approximately 3~\cite{Cheng2019}.

\subsection{Disease-Gene Associations}
Disease-gene associations were obtained from the BioSNAP 'DG-AssocMiner' dataset (Stanford SNAP Biomedical Network Dataset Collection) \cite{biosnapnets}, downloaded from \url{https://snap.stanford.edu/biodata/datasets/10012/10012-DG-AssocMiner.html} on 4 October 2024. The dataset contains 21,357 associations from 519 diseases to 7,294 gene nodes.

To define disease modules within the human interactome, we mapped these gene nodes onto the aforementioned interaction network using their respective Entrez IDs. Only genes whose protein products appear as nodes in $\mathcal{G}$ were retained, since the network proximity measures employed in this work ($z$-score and the separation $s$) are defined in terms of shortest-path distances on $\mathcal{G}$ and are therefore undefined for proteins outside its node set. The disease module for disease $Y$ is thus the subset $Y \subset V$ resulting from this intersection. We do not require $Y$ to form a single connected component, and all shortest-path computations are performed on the full graph $\mathcal{G}$ regardless of the internal connectivity of $Y$.

Allowing for fragmented modules is topologically and biologically justified by the disease module hypothesis~\cite{Barabasi2011}, which posits that disease-associated proteins agglomerate into localized network neighborhoods rather than distributing randomly. Menche \textit{et al.}~\cite{Menche2015} demonstrated this quantitatively across hundreds of diseases, showing that the largest connected subgraph formed by disease proteins ($S_i$) is typically much larger than expected by chance, and that the mean shortest distance to the nearest same-disease neighbor ($\langle d_s \rangle$) is significantly reduced. Crucially, even proteins that remain disconnected from $S_i$, which can constitute up to 80\% of known disease genes due to the inherent incompleteness of the human interactome, still predominantly localize in the immediate network vicinity of the observable module~\cite{Menche2015}. Because of this residual localization, shortest-path-based proximity measures maintain a robust and meaningful biological signal even when the underlying disease module is highly fragmented.

\subsection{Validated Drug Combinations}
The disease-specific validated Drug Combinations dataset was constructed by intersecting the multi-drug combinations reported in the Continuous Drug Combination Database (CDCB) \cite{shtar2022cdcdb} with drug-disease associations compiled by Guney \textit{et al.} \cite{Guney2016}. A disease-combination pair was retained only if (i) the combination appears in CDCB as a verified multi-drug combination, and (ii) every drug in that combination is associated with the disease in Guney \textit{et al.} . That is, we required exact agreement between the set of drugs in the combination and the set of drugs associated with the disease. Applying this overlap and selection rule yields disease-combination pairs spanning 35 distinct diseases with at least one combination, 136 unique drugs, and 287 unique disease-combination pairs, 234 of which are constituted by two drugs, 40 are combinations of three drugs, and 13 of four or more drugs. The resulting dataset is publicly available in the associated project repository: \url{https://github.com/dmrapk/Drug-Combinations-using-Quantum-Annealers}.

\section{Literature Validation of Predicted Combinations}
\label{app:literature_validation}

To assess whether the low-energy configurations identified by the Hamiltonian encode biologically meaningful information beyond the validated dataset $\mathcal{C}$, we conducted a modest examination of the novel predictions (those not present in $\mathcal{C}$) against the published clinical literature. 
We stress that this analysis is not a substitute for a proper pharmacological evaluation, which would require domain expertise in medicinal chemistry and clinical pharmacology to assess safety, dosing, and translational feasibility. 
Our goal is limited to checking whether independently published evidence exists for any of the predicted pairings.

The most compelling case is the predicted combination of adalimumab and methotrexate for Rheumatoid Arthritis (energy rank $3$ in the enlarged candidate set). 
This pairing was independently evaluated in the PREMIER trial~\cite{breedveld2006premier}, a randomised, double-blind study of $799$ patients. 
After two years, patients receiving the combination exhibited a clinical remission rate approximately double that of patients receiving either drug alone~\cite{breedveld2006premier}. 
The algorithm identifies this combination as one of the lowest-energy configurations without access to any clinical outcome data, relying solely on the topological relationship between the two drugs' protein targets and the disease module in the interactome.

For Diabetes Mellitus, the top ranked novel prediction pairs metformin with irbesartan. 
Kidney damage is one of the most common long-term complications of diabetes, and Ishibashi et al.~\cite{ishibashi2012beneficial} studied this exact combination in a laboratory setting, finding that the two drugs protect kidney cells from diabetes-related damage through independent and additive mechanisms. 
While this evidence is preclinical rather than clinical, it is notable that the study explicitly investigates the same drug pair that our algorithm identifies.

Concerning Asthma, several of the highest ranked predicted combinations involve arformoterol, a bronchodilator that is approved for chronic obstructive pulmonary disease (COPD) rather than asthma \cite{fda_brovana_2019}. 
Its appearance among the lowest-energy configurations may reflect a likely considerable degree of network proximity between the COPD and asthma disease modules in the interactome.

In the case of Brain Neoplasms, a brief literature search reveals that several of the recurring drugs among the top predictions (vinorelbine, epirubicin, tamoxifen, and capecitabine) appear predominantly in the context of breast cancer treatment rather than brain tumours. 
This pattern could indicate that the algorithm is capturing shared oncogenic network signatures rather than brain-specific therapeutic relevance, which would be consistent with the lower AP obtained for this disease (Table~\ref{tab:hyperparams}).

We emphasize that these predictions reflect topological proximity within the interactome rather than guaranteed biological efficacy.
A comprehensive evaluation by specialists in the relevant therapeutic areas would be needed to assess their clinical viability. 
An extended table of the top predictions for all four diseases, including literature references where available, is provided in the project repository at \url{https://github.com/dmrapk/Drug-Combinations-using-Quantum-Annealers}.

\end{document}